\documentclass[eqsecnum,amsmath,amsfonts,amssymb,prd,a4paper,12pt]{revtex4}
\usepackage{graphicx}
\usepackage{color}
\usepackage{psfrag}
\usepackage{psfig}
\usepackage{epsfig}
\usepackage{ifthen}
\usepackage{calc}
\usepackage{graphpap}
\usepackage{supertabular}
\usepackage{longtable}
\usepackage{bm}
\newcommand{\topbott}[2]{\left\{ #1 \atop #2 \right\}}
\renewcommand{\d}{\mathrm{d}}
\newcommand{\diff}[2]{
  \ifthenelse{\equal{#1}{}}%
{\frac{\mathrm{d}\hphantom{#2}}{\mathrm{d}#2}}%
{\frac{\mathrm{d}#1}{\mathrm{d}#2}}}
\newcommand{\ddiff}[2]{
  \ifthenelse{\equal{#1}{}}%
{\frac{\mathrm{d}^2\hphantom{#2}}{\mathrm{d}#2^2}}
{\frac{\mathrm{d}^2#1}{\mathrm{d}#2^2}}}
\newcommand{\pardiff}[2]{\frac{\partial#1}{\partial#2}}
\newcommand{\parddiff}[2]{\frac{\partial^2#1}{\partial#2^2}}
\newcommand{\pparddiff}[3]{\frac{\partial^2#1}{\partial#2\,\partial#3}}

\newcommand{\indhel}{h}

\begin{document}

\title{\textbf{
High Frequency Asymptotics for\\ the Spin-Weighted Spheroidal
Equation }}

\author{Marc Casals}
\email{marc.casals@ucd.ie}
\author{Adrian C. Ottewill}
\email{adrian.ottewill@ucd.ie}
 \affiliation{Department of Mathematical Physics,University
College Dublin, Belfield, Dublin 4, Ireland}

\begin{abstract}
We fully determine a uniformly valid asymptotic behaviour for large $a \omega$
and fixed $m$ of the angular solutions and eigenvalues of the
spin-weighted spheroidal differential equation. We fully
complement the analytic work with a numerical study.
\end{abstract}

\pacs{????}
\maketitle


\section{Introduction} \label{sec:Introduction}

By making use of the Newman-Penrose formalism, Teukolsky
(~\cite{ar:Teuk'72}, ~\cite{ar:Teuk'73}) showed that the equations
describing linear scalar (spin-0), neutrino (spin-1/2)
electromagnetic (spin-1), and gravitational (spin-2) perturbations
of a general Type D background can be decoupled.  Using the
Kinnersley tetrad and Boyer-Lindquist co-ordinates, Teukolsky
wrote the field equations in the Kerr background in compact form
for the various spin fields, as one single `master' equation
\begin{equation}
\label{eq:Teuk.eq.}
\begin{aligned}
&\left[\frac{(r^2+a^2)^2}{\Delta}-a^2\sin^2\theta\right]\parddiff{\Omega_h}{t}+
\frac{4Mar}{\Delta}\pparddiff{\Omega_h}{t}{\phi}+
\left[\frac{a^2}{\Delta}-\frac{1}{\sin^2\theta}\right]\parddiff{\Omega_h}{\phi}- \\
&-\Delta^{-h}\frac{\partial}{\partial
r}\left(\Delta^{h+1}\pardiff{\Omega_h}{r}\right)-
\frac{1}{\sin\theta}\frac{\partial}{\partial
\theta}\left(\sin\theta\pardiff{\Omega_h}{\theta}\right)-
2h\left[\frac{a(r-M)}{\Delta}+\frac{i\cos\theta}{\sin^2\theta}\right]\pardiff{\Omega_h}{\phi}- \\
&-2h\left[\frac{M(r^2-a^2)}{\Delta}-r-ia\cos\theta\right]\pardiff{\Omega_h}{t}+(h^2\cos^2\theta-h)\Omega_h=
4\pi\Sigma T_h
\end{aligned}
\end{equation}
where $\Delta = r^2-2Mr +a^2$ and $\Sigma=r^2+a^2\cos^2\theta$.
The field $\Omega_h$ and the source term $T_h$ are defined in ~\cite{ar:Teuk'73}. 
The parameter $\indhel=0,\pm 1/2, \pm1,\pm2$ refers to the helicity of the field.
Following on Carter's work~\cite{ar:Carter'68b} for the scalar
(spin-0) case, Teukolsky further showed that in the Kerr
background the homogeneous decoupled equations can be
solved by separation of variables:
\[
{}_{lmw}\Omega_h(t,r,\theta,\phi) \propto
            {}_{h}R_{lm\omega}(r){}_{h}S_{lm\omega}(\theta)e^{-i\omega t}e^{+im\phi}
\]
The angular equation resulting from the separation of the
Teukolsky equation is the so-called spin-weighted spheroidal
differential equation and its  regular solutions are the
spin-weighted spheroidal harmonics (SWSH). In terms of
$x=\cos\theta$, the spin-weighted spheroidal differential equation
is
\begin{equation} \label{eq:ang. teuk. eq.}
\left[
\frac{\mathrm{d}\phantom{x}}{\mathrm{d}x}\left((1-x^{2})
\frac{\mathrm{d}\phantom{x}}{\mathrm{d}x}\right)+c^{2}x^{2}-2\indhel
cx-\frac{(m+\indhel
x)^{2}}{1-x^{2}}+{}_{\indhel}E_{lm\omega}-\indhel^2 \right]
{}_{\indhel}S_{lm\omega}(x)=0
\end{equation}
where ${}_{\indhel}E_{lm\omega}$ denotes the eigenvalue and $c=a\omega$. It would be more logical to label the angular solutions and the
eigenvalues by $c$ rather than $\omega$, but following
convention we label them by $\omega$. The eigenvalue for the case
$c=0$, corresponding to  Schwarzschild space-time, is well-known to be
\begin{equation} \label{eq:eigenval. for c=0}
{}_{\indhel}E_{lm\omega=0}=
l(l+1)-\indhel (\indhel+1),
\end{equation}
with regular solutions being the spin-weighted spherical harmonics~\cite{ar:Gold.etal.}.

The corresponding radial equation is
\begin{equation} \label{eq:radial teuk. eq.}
\Delta^{-\indhel }\diff{}{r}\left(\Delta^{\indhel+1}\diff{{}_{\indhel}R_{lm\omega}}{r}\right)-{}_{\indhel}V{}_{\indhel}R_{lm\omega}=0
\end{equation}
where the potential is given by
\begin{equation} \label{eq:radial teuk. potential}
{}_{\indhel}V=\frac{2i\indhel (r-M)K-K^2}{\Delta}-4i\indhel \omega r+{}_{\indhel}\lambda_{lm\omega}
\end{equation}
with $K=(r^2+a^2)\omega -am$. The separation constants are related by
\begin{equation} \label{eq:lambda}
{}_{\indhel}\lambda_{lm\omega} \equiv
{}_{\indhel}E_{lm\omega}-\indhel (\indhel+1)+c^{2}-2mc
\end{equation}

The differential equation (\ref{eq:ang. teuk. eq.}) has two
regular singular points at $x=\pm 1$ and one essential singularity
at $x=\infty$.
We are only interested in solutions for real values of
the independent variable $\theta$ corresponding to the interval
$x\in[-1,+1]$. 
We henceforth restrict $x$ to this range and therefore we have
only to consider the two regular singular points at $x=\pm 1$. The
differential equation (\ref{eq:ang. teuk. eq.}), together with the
boundary condition that its solution ${}_{\indhel}S_{lm\omega}(x)$
is regular for $x\in[-1,+1]$,
defines a parametric eigenvalue problem, with parameters $c$, $m$ and $\indhel$.
The physical requirements of single-valuedness and
of regularity at $x=\pm 1$ requires that $l$ and $m$ are integers
with $|m|\leq l$. 
Stewart
~\cite{ar:Stewart'75} showed that the SWSH form a strongly
complete set if $c$ is real while he could only prove weak
completeness if $c$ is complex.

In this paper we study the asymptotic behaviour for high frequency of the solution
and eigenvalues of the spin-weighted spheroidal differential equation.
Following standard conventions, we refer to
`high frequency' in relation to the angular solution and eigenvalues when
in fact what it is meant is large $c(=a\omega$), where $a$ is the angular momentum per unit mass of the rotating black hole
and $\omega$ is the frequency of the mode.

The high frequency approximation of the spin-weighted spheroidal
equation is a particularly important subject that has
been left unresolved thus far, except for the spin-0 case, due to its difficulty.
This asymptotic study is
important when considering both classical and quantum perturbations. In the
classical case it is important, for example, when calculating
gravitational radiation emitted by a particle near the black hole
since the typical time-scale of the motion is short compared to the
scale set by the curvature of the black hole. In the quantum
case its importance lies in the fact that the high frequency
limit is at the root of the divergences that the expectation value
of the stress-energy tensor possesses. The correct subtraction of
the divergent terms from the expectation value of the stress-energy
tensor is extremely troublesome in curved space-time,
particularly in one that is not spherically symmetric.
As the divergent terms arise from the high
frequency behaviour of the field, knowledge of this
behaviour is fundamental in such a subtraction.
This limit has also been recently considered in the Kerr background in the context of
quasinormal modes (see ~\cite{ar:Berti&Card&Yosh'04}).

All
analysis in this paper
has been performed for general spin, so that it applies to the scalar, neutrino, electromagnetic
and, in particular, gravitational perturbations, which are of great interest in astrophysics.
However, we should note that the asymptotic study in this
paper is valid for fixed $m$ as $c$ tends to
infinity, a fuller understanding of the asymptotic behaviour of
the solution would require an anlysis uniform in $m$.

In the remainder of this introductory section we discuss the
results for high frequency asymptotics of SWSH that have been
obtained in the literature up until now, show their shortcomings
and outline what our new results achieve. In the next section we
lay down the basic theory that we use in the following sections. In
Sections \ref{sec:Inner solutions}, \ref{sec:Outer solution},
\ref{sec:Matching the solutions} and \ref{sec:Evaluation of
gamma} we fully determine the aymptotic behaviour of the angular solution that is uniform in $x$
and the asymptotic behaviour of the eigenvalue. In
Section \ref{sec:num. method; high freq. sph.} we describe the
numerical method and programs used to obtain the numerical results,
which in the last section we show, analyze and compare to
our asymptotic results and to numerical results in the literature.

Different authors have obtained high-frequency approximations to the solution and eigenvalues of the spheroidal
differential equation, which results from the spin-weighted spheroidal differential equation when $\indhel=0$.
Erd\'{e}lyi et al. ~\cite{bk:high_transc_funcs}, Flammer ~\cite{bk:Flammer} and Meixner and Sch\"{a}fke
~\cite{bk:Meixner&Schafke} have all done so using the fact that the spheroidal differential equation
becomes the Laguerre differential equation in the high-frequency limit.

Breuer ~\cite{ar:Breuer'75} was the first author to study the
high-frequency behaviour of the spin-weighted spheroidal
harmonics. Based on the work on the spin-0 case by the above
authors  he related the solution of a transformation of the
spin-weighted spheroidal equation for large $c$ and finite $m$ to
generalized Laguerre polynomials. His work, however, was
fundamentally flawed as it assumed that the solution was
either symmetric or antisymmetric under $\theta\rightarrow \pi-\theta$, which is only true for spin-0.

Breuer, Ryan and Waller ~\cite{ar:BRW} (hereafter referred to as BRW) corrected this error and further developed this study by first
relating the SWSH to the confluent hypergeometric functions and then
reducing them to the generalized Laguerre polynomials
by imposing regularity far from the boundary points $x=\pm 1$, where $x\equiv\cos\theta$. 
Unfortunately, their study of the high-frequency behaviour
was also flawed and incomplete. The behaviour for high frequency of
both the spherical functions and the eigenvalues obtained by BRW
depend critically on a certain parameter $\gamma$ (called $\alpha$
in that paper) which they were unable to determine for the case of non-zero spin.

BRW obtained the analytic value of $\gamma$ for the spin-0 case, however for non-zero spin they could
only calculate it numerically for a handful of sets of values of
$\{l,m\}$ for spin-2. BRW achieved this numerical calculation for the spin-2 case
by matching the high-frequency asymptotic expression for the eigenvalue that they obtained with the
expression for the eigenvalue given by Press and Teukolsky
~\cite{ar:Press&Teuk'73} valid for low frequency. Not only their
analytic expressions for both the spherical solution and the
eigenvalue for high frequency were thus left undetermined, but
also their expressions for the spherical solution are only valid
sufficiently close to the boundary points $x=1$ and $x=-1$,
not for the region in-between them. This results in the
possibility that a zero of the solution near $x=0$, away
from $x=\pm1$, be overlooked. Furthermore, and crucially, their assumption that
the confluent hypergeometric functions should reduce to the
generalized Laguerre polynomials by imposing regularity far from
the boundary points is not correct. The reason why it is not correct is that in the
cases for which the confluent hypergeometric function diverges far
from one of the boundaries, the coefficient in front of it
decreases exponentially with $c$ so that the solution remains finite
in the whole region $x\in[-1,+1]$.  
We believe that the reason why they were not able to
analytically determine the value of the parameter $\gamma$
is because they ignored the behaviour of the
solution far from the boundaries, thus overlooking a possible
zero, and wrongly imposed regularity.

The study of the behaviour of the solution and eigenvalues of the
spin-weighted spheroidal equation for high frequency and finite $m$ has not been
developed any further by these or any other authors and
therefore BRW's work is where this study stood until the present paper.

In this paper we correct and complete BRW's study for high
frequency and finite $m$. We thus obtain an asymptotic solution for large
frequency to the spin-weighted spheroidal equation which is
uniformly valid everywhere within the range $x\in[-1,+1]$, not just near the boundaries.
We also analyze the existence and location of a possible zero of the solution near $x=0$.
We analytically determine the value of $\gamma$ by matching the
number of zeros that our asymptotic solution has with the number of zeros the SWSH has.
As a consequence, the
asymptotics of the eigenvalue in the same limit also become fully determined.
Finally, we have complemented all the analytic work with graphs
produced with data that we obtained numerically. The graphs show the
behaviour of the eigenvalues for large frequency and how they
match with Press and Teukolsky's approximation for low frequency.
They also show the behaviour of the SWSH in this limit and the
location of its zeros.


\section{Symmetries of the spin-weighted spheroidal differential equation} \label{sec:SWSH}

Certain symmetries of the spin-weighted spheroidal equation~(\ref{eq:ang. teuk. eq.})
are immediate: the equation remains invariant under the change
in sign of two quantities among $[s, (m,\omega), x]$, where we are considering that $(m,\omega)$
constitutes one single quantity, i.e., $m$ and $\omega$ change sign simultaneously.       
As a consequence, the SWSH satisfy the following symmetries, where the choice of signs
ensures consistency with the Teukolsky-Starobinski\u{\i} identities below:
\begin{subequations} \label{eq: S symms}
\begin{align}
&{}_{\indhel}S_{lm\omega}(\theta)=(-1)^{l+m}{}_{-\indhel}S_{lm\omega}(\pi-\theta)   \label{eq:S symm.->pi-t,-s} \\
&{}_{\indhel}S_{lm\omega}(\theta)=(-1)^{l+\indhel }{}_{\indhel}S_{l-m-\omega}(\pi-\theta)  \label{eq:S symm.->pi-t,-m,-w}\\
&{}_{\indhel}S_{lm\omega}(\theta)=(-1)^{\indhel+m}{}_{-\indhel}S_{l-m-\omega}(\theta)     \label{eq:S symm.->-s,-m,-w}
\end{align}
\end{subequations}
Here any one symmetry follows from the other two.
The eigenvalues must consequently also satisfy the symmetries:
\begin{subequations} \label{eq:eigenval. symms.}
\begin{align}
&{}_{\indhel}E_{lm\omega}={}_{-\indhel}E_{lm\omega}   \label{eq:eigenval. symm.->-s} \\
&{}_{\indhel}E_{lm\omega}={}_{\indhel}E_{l-m-\omega}  \label{eq:eigenval. symm.->-m,-w}
\end{align}
\end{subequations}

The SWSH with helicity $\indhel$ is related to the SWSH with helicity $-\indhel$
via the Teukolsky-Starobinski\u{\i} identities ~\cite{ar:Teuk&Press'74}.
We start by defining the operator
\begin{equation} \label{eq:def. L_n}
\mathcal{L}_{n}^
{ \topbott{}{\dagger}} \equiv \partial_{\theta} \mp c \sin\theta \pm \frac{m}{\sin\theta}+n\cot\theta .
\end{equation}
Then for spin-$\frac12$, the Teukolsky-Starobinski\u{\i} identities may be written as
\begin{subequations} \label{eq:Teuk-Starob. ids.s=1/2}
\begin{align}
\mathcal{L}_{\frac12}{}_{+{\frac12}}S_{lm\omega}&=-{}_{\frac12}\mathcal{B}_{lm\omega}{}_{-{\frac12}}S_{lm\omega}  \\
\mathcal{L}^{\dagger}_{\frac12}{}_{-{\frac12}}S_{lm\omega}&={}_{\frac12}\mathcal{B}_{lm\omega}{}_{+{\frac12}}S_{lm\omega}
\end{align}
\end{subequations}
where
\begin{equation} 
{}_{\frac12}\mathcal{B}_{lm\omega}^2= {}_{-{\frac12}}\lambda_{lm\omega} .
\end{equation}
For the spin-1 case, the Teukolsky-Starobinski\u{\i} identities may be written as
\begin{subequations} \label{eq:Teuk-Starob. ids.}
\begin{align}
\mathcal{L}_0\mathcal{L}_1{}_{+1}S_{lm\omega}&={}_1\mathcal{B}_{lm\omega}{}_{-1}S_{lm\omega}  \\
\mathcal{L}^{\dagger}_0\mathcal{L}^{\dagger}_1{}_{-1}S_{lm\omega}&={}_1\mathcal{B}_{lm\omega}{}_{+1}S_{lm\omega}
\end{align}
\end{subequations}
where
\begin{equation} \label{eq:def. B}
{}_1\mathcal{B}_{lm\omega}^2= {}_{-1}\lambda_{lm\omega}^2+4mc-4c^2
\end{equation}
Finally for the spin-2 case, the Teukolsky-Starobinski\u{\i} identities may be written as
\begin{subequations} \label{eq:Teuk-Starob ids. for spher.s=2}
\begin{align}
\mathcal{L}_{-1}\mathcal{L}_{0}\mathcal{L}_{1}\mathcal{L}_{2} {}_{+2}S_{lm\omega}&={}_{2}\mathcal{B}_{lm\omega} {}_{-2}S_{lm\omega} \\
\mathcal{L}^{\dagger}_{-1}\mathcal{L}^{\dagger}_{0}\mathcal{L}^{\dagger}_{1}\mathcal{L}^{\dagger}_{2} {}_{-2}S_{lm\omega}&={}_{2}\mathcal{B}_{lm\omega} {}_{+2}S_{lm\omega}
\end{align}
\end{subequations}
where
\begin{equation}
\begin{aligned}
{}_{2}\mathcal{B}_{lm\omega}^{2}&=
{}_{-2}\lambda_{lm\omega}^{2}({}_{-2}\lambda_{lm\omega}+2)^{2}
-8c^{2}{}_{-2}\lambda_{lm\omega}\left\{\left(1-\frac{m}{c}\right)\left[5{}_{-2}\lambda_{lm\omega}+6\right]-12\right\}+ \\
&+144c^{4}\left(1-\frac{m}{c}\right)^{2}
\end{aligned}
\end{equation}
The signs of ${}_{\frac12} \mathcal{B}_{lm\omega}$, ${}_1\mathcal{B}_{lm\omega}$ and ${}_2\mathcal{B}_{lm\omega}$ are arbitrary,
but we will take them to be both positive. With this convention,
(\ref{eq:Teuk-Starob. ids.s=1/2}), (\ref{eq:Teuk-Starob. ids.}) and (\ref{eq:Teuk-Starob ids. for spher.s=2}) agree with
the sign in the 
symmetry (\ref{eq:S symm.->pi-t,-s}) of the angular function.

\section{The Uniform asymptotic solution}

In the rest of this paper we follow the approach to boundary
layer theory as presented by Bender and
Orszag~\cite{bk:Bender&Orszag}. The asymptotic solution that is
a valid approximation to the solution of the differential equation
from the boundary point $\pm 1$ until $x\sim \pm 1+O(c^{\delta})$,
where $-1\leq\delta<0$, is called the
inner solution. The region within which an inner solution
is valid is a
boundary layer. As we shall see, for the large frequency
approximation of the spin-weighted spheroidal equation, there are
two boundary layers within the region $x\in[-1,1]$, one close to
$x=-1$ and one close to $x=+1$. Close to the boundary points the
SWSH oscillate rather quickly in
$x$, and indeed it is there where all the zeros of the function are located 
(with the possible exception of one).

The asymptotic solution that is a valid approximation to the solution of the differential equation
in the range $-1+O(c^{-1})\ll x \ll +1-O(c^{-1})$, is called
the outer solution. This range comprises not only the region in between the two boundary layers but also a certain region
of both boundary layers. This region where both an inner solution and the outer solution are valid is the
overlap region, and it is there that the outer and inner solutions are matched.

We shall see that in between the two boundary layers the function behaves rather smoothly, like a $\cosh x$
or a $\sinh x$, 
so that the SWSH may have at the most one zero, which will turn out to lie close to $x=0$.
The behaviour of the outer solution is important despite its
smoothness because when matching it with the inner solutions it
will allow us to find an asymptotic solution which is uniformly
valid throughout the whole range of $x$. The outer solution is
also necessary in order to find out whether or not the uniform
solution has a zero close to $x=0$ and, if it does, to calculate
the analytic location of the zero.

This is a key feature that singles out the scalar case from the others: for the spin-$0$ case the differential equation
(\ref{eq:ang. teuk. eq.}) is clearly symmetric under $\{ x\leftrightarrow -x\}$ and therefore, depending
on its parity, it will have a zero at $x=0$ or not. On the other hand, for the case of spin non-zero, the differential equation does not
satisfy this symmetry but it does remain unchanged under the transformation
$\left \{ x\leftrightarrow -x, \indhel \leftrightarrow -\indhel \right \}$
instead. There is therefore no apparent reason why it should have
a zero near the origin. The outer solution is important for the
case of spin non-zero and not for spin-$0$ since, as we shall see, the
differential equation that the outer solution satisfies is symmetric under
$\left \{ x\leftrightarrow -x, \indhel \leftrightarrow -\indhel \right \}$
to leading order in $c$.

As noted above, the differential equation (\ref{eq:ang. teuk.
eq.}) has singular points
at $x=\pm 1$. 
By using the Frobenius method it can be found that the solution
that is regular at both
boundary points $x=+1$ and $-1$ is given by                             
\begin{equation} \label{eq:asympt. S for x->+/-1}
{}_{\indhel}S_{lm\omega}(x)=(1-x)^{\alpha}(1+x)^{\beta}{}_{\indhel}y_{lm\omega}(x)
\end{equation}
where
\begin{equation}
\begin{aligned}
\alpha &=\frac{|m+\indhel |}{2}, & \qquad \qquad
\beta&=\frac{|m-\indhel |}{2}
\end{aligned}
\end{equation}
and the function ${}_{\indhel}y_{lm\omega}(x)$ satisfies the
differential equation
\begin{equation} \label{eq:ang. teuk. eq. for y}
\begin{aligned}
& \Bigg\{
 (1-x^{2})\ddiff{}{x}-2\left[\alpha-\beta+(\alpha+\beta+1)x\right]\diff{}{x}+
\\ &
\qquad \qquad \qquad \qquad \qquad \qquad
+{}_{\indhel}E_{lm\omega}-(\alpha+\beta)(\alpha+\beta+1)+c^{2}x^{2}-2\indhel
cx \Bigg\} {}_{\indhel}y_{lm\omega}(x)=0
\end{aligned}
\end{equation}

\subsection{Inner solutions} \label{sec:Inner solutions}

BRW obtained an expression for the inner solution for general spin in terms of an undetermined parameter $\gamma$.
In this section we summarize and present their results in a compact way.

By making the variable substitution $u=2c(1-x)$, equation (\ref{eq:ang. teuk. eq. for y}) becomes
\begin{equation} \label{diff eq:y in u}
\begin{aligned}
&u\frac{d^{2}{}_{\indhel}y_{lm\omega}}{du^{2}}+(2\alpha+1)\frac{d{}_{\indhel}y_{lm\omega}}{du}-  \\
&\quad-\frac{1}{4}\left\{u+2\indhel -\frac{1}{c}\left[c^{2}-(\alpha+\beta)(\alpha+\beta+1)+{}_{\indhel}E_{lm\omega}\right]
\right\}{}_{\indhel}y_{lm\omega}-
\\
&\quad-\frac{1}{4c}\left\{u^{2}\frac{d^{2}{}_{\indhel}y_{lm\omega}}{du^{2}}+2(\alpha+\beta+1)\frac{d{}_{\indhel}y_{lm\omega}}{du}-
\left(\frac{1}{4}u^{2}+\indhel u\right){}_{\indhel}y_{lm\omega}\right\}=0
\end{aligned}
\end{equation}
It is clear from this equation that the leading
order behaviour of ${}_{\indhel}E_{lm\omega}$ for large $c$ must
be:
\begin{equation} \label{eq:series E for large w}
{}_{\indhel}E_{lm\omega}=-c^{2}+\gamma c+O(1).
\end{equation}
If its leading order were not $-c^2$, there would then be a leading
order term $+\frac{1}{4}c{}_{\indhel}y_{lm\omega}$ in the equation that it
could not be matched with any other term. Lower order
terms for ${}_{\indhel}E_{lm\omega}$ are given in BRW. It is crucial to know
the value of the parameter $\gamma$, as it determines how the
angular function behaves asymptotically to leading order in $c$.
At this stage, $\gamma$ is an undetermined real number;
we will determine its value later on.

Using the asymptotic behaviour (\ref{eq:series E for large w}) and letting $c\to \infty$, the terms in
(\ref{diff eq:y in u}) of order $O(c^{-1})$ can be ignored
with respect to the other ones and, to leading order in $c$, the function ${}_{\indhel}y_{lm\omega}$ satisfies
\begin{equation} \label{diff eq:y in u,1st order}
u\frac{d^{2}{}_{\indhel}y_{lm\omega}}{du^{2}}+(2\alpha+1)\frac{d{}_{\indhel}y_{lm\omega}}{du}-\frac{1}{4}\left(u+2\indhel -\gamma \right){}_{\indhel}y_{lm\omega}=0.
\end{equation}
The solution of this differential equation that satisfies the boundary
condition of regularity at $x=+1$ is related to the confluent
hypergeometric function:
\begin{equation}
{}_{\indhel}y_{lm\omega}^{\text{inn},+1}={}_{\indhel}C_{lm\omega}e^{-u/2}{}_1F_{1}\Big((|m+\indhel |+\indhel+1)/2-\gamma/4,|m+\indhel |+1,u\Big)
\end{equation}
where ${}_{\indhel}C_{lm\omega}$ is a constant of integration.

Similarly, if we instead make a change of variable $u^{*}=2c(1+x)$ in equation (\ref{eq:ang. teuk. eq. for y}), due to the
$\left\{ x\leftrightarrow -x, \indhel\leftrightarrow -\indhel  \right \}$ symmetry we obtain
\begin{equation}
{}_{\indhel}y_{lm\omega}^{\text{inn},-1}={}_{\indhel}D_{lm\omega}e^{-u^{*}/2}{}_1F_{1}\Big((|m-\indhel |-\indhel+1)/2-\gamma/4,|m-\indhel |+1,u^{*}\Big)
\end{equation}
as the solution that is regular at $x=-1$.

We use the following obvious notation to refer to the solutions of the spin-weighted spheroidal
equation that correspond to the inner solutions of (\ref{diff eq:y in u,1st order}):
\[
{}_{\indhel}S_{lm\omega}^{\text{inn},\pm 1}=(1-x)^{\alpha}(1+x)^{\beta}{}_{\indhel}y_{lm\omega}^{\text{inn},\pm 1}
\]
The inner solution ${}_{\indhel}S_{lm\omega}^{\text{inn},\pm 1}$ is only a valid
approximation in the region from the boundary point $\pm 1$ until $\pm 1-x\sim
O(c^{\delta})$ with $-1\leq\delta<0$. The reason is that in the
step from (\ref{diff eq:y in u}) to (\ref{diff eq:y in u,1st
order}) we have ignored terms with $u^{\topbott{}{*}}/c$ with respect to terms of order
$O(1)$, and therefore the inner solution has been found for
$\pm 1 -x \sim u^{\topbott{}{*}}/c\ll O(1)$
and so we must have $\delta<0$. On the other hand, we are not
ignoring $u$ with respect to the $O(1)$ term $(2\indhel -\gamma)$ in
equation (\ref{diff eq:y in u,1st order}), so that it must be
$u\sim O(c^{\delta+1})$ with $\delta+1\geq 0$.
From the fact that we are not ignoring $(2\indhel -\gamma)$ with respect to $u$
it does not follow that $\delta+1\leq 0$, since the inner solution is valid at
the boundary point $x=+1$, where $u=0$.
That is, the term $(2\indhel -\gamma)$ cannot be ignored with respect to $u$
for all $x$ from $+1$ up to $+1-x\sim O(c^{\delta})$, even if $\delta+1\geq 0$.
A similar reasoning applies to $u^*$.

We therefore have one boundary layer comprising the region in $x$ from
$-1$ to  $(-1-x)\sim O(c^{\delta})$ and another boundary layer
from $(+1-x)\sim O(c^{\delta})$ to $+1$.

To leading order in $c$ the solution to the
spin-weighted spheroidal equation which is valid within the two boundary layers is given by
\begin{equation} \label{eq: inner solution}
{}_{\indhel}S_{lm\omega}^{\text{inn}} =(1-x)^{\alpha}(1+x)^{\beta}
\begin{cases}
{}_{\indhel}C_{lm\omega}e^{-u/2}{}_1F_{1}(-p,2\alpha+1,u)         & \qquad x>0 \\
{}_{\indhel}D_{lm\omega}e^{-u^{*}/2}{}_1F_{1}(-p',2\beta+1,u^{*}) & \qquad x<0
\end{cases}
\end{equation}
where we have defined
\begin{equation} \label{eq:def pp'}
\begin{cases}
p\equiv -(|m+\indhel |+\indhel+1)/2+\gamma/4    \\
p'\equiv -(|m-\indhel |-\indhel+1)/2+\gamma/4
\end{cases}
\end{equation}

BRW then require that $p,p'\in \mathbb{Z}^{+}$ in order that
the inner solution ${}_{\indhel}S_{lm\omega}^{\text{inn}}$ is regular at $x=0$, where
$u, u^*\rightarrow \infty$.
Correspondingly, they replace the
confluent hypergeometric functions ${}_1F_{1}(a,b,x)$ by the
generalized Laguerre polynomials $L_{-a}^{(b-1)}(x)$.
As we shall see, this is erroneous:
$p,p'\in \mathbb{Z}^{+}$ is not a necessary condition for
regularity since in the cases for which this condition is not
satisfied, the coefficients ${}_{\indhel}C_{lm\omega}$ and ${}_{\indhel}D_{lm\omega}$
diminish exponentially for large $c$ in such a way that
${}_{\indhel}S_{lm\omega}^{\text{inn}}$ remains regular.


\subsection{Outer solution} \label{sec:Outer solution}

We now proceed to find the outer solution of the spin-weighted spheroidal differential equation.
We first make the variable substitution
\begin{equation}
y(x)=g(x)\exp{\int
\frac{\alpha-\beta+(\alpha+\beta+1)x}{1-x^2}dx}=
g(x)(1-x)^{-(2\alpha+1)/2}(1+x)^{-(2\beta+1)/2}
\end{equation}
which transforms equation (\ref{eq:ang. teuk. eq.}) into
\begin{equation} \label{eq:g}
g''(x)+f(x,c)g(x)=0
\end{equation}
where
\begin{equation}
\begin{aligned}
f(x,c)&=\frac{G(x,c)}{1-x^{2}}+
\frac{(\alpha+\beta+1)(1-x^{2})+2x\left[\alpha-\beta+(\alpha+\beta+1)x\right]}{(1-x^{2})^{2}}-
\\&
-\frac{\left[\alpha-\beta+(\alpha+\beta+1)x\right]^{2}}{(1-x^{2})^{2}}
\end{aligned}
\end{equation}
and $G(x,c)$ is the coefficient of ${}_{\indhel}y_{lm\omega}$ in (\ref{eq:ang. teuk. eq. for y}), i.e.,
\begin{equation}
G(x,c)={}_{\indhel}E_{lm\omega}-(\alpha+\beta)(\alpha+\beta+1)+c^{2}x^{2}-2\indhel cx
\end{equation}

We now perform a WKB-type expansion: $g(x)=e^{\mathcal{G}(x)}$. This change of variable converts equation (\ref{eq:g}) into
\begin{equation} \label{eq:phi}
\mathcal{G}''(x)+\mathcal{G}'(x)^{2}+f(x,c)=0
\end{equation}
Performing an asymptotic expansion of $f(x,c)$ in $c$ we find
\begin{equation}
f(x,c)=f_{0}(x)c^{2}+f_{1}(x)c+O(1),
\end{equation}
with
\begin{equation}
f_{0}(x)=-1, \qquad f_{1}(x)=\frac{2(q-\indhel x)}{1-x^{2}} ,
\end{equation}
where we have used the asymptotic expansion of ${}_{\indhel}E_{lm\omega}$ in
$c$ and we have also introduced the parameter $q \equiv
\gamma/2$. We will prove in Section~\ref{sec:Evaluation of gamma} that $q$ must be an integer.
It is clear that to leading order in $c$ the outer solution is
symmetric under $\{ x\leftrightarrow -x\}$.
We are avoiding any possible turning points 
by assuming that $f(x,c)\neq 0$ for $x$ values of interest.
This condition is clearly satisfied if $c$ is large enough.

Next we perform an asymptotic expansion of $\mathcal{G}(x)$ in $c$.
We do not know a priori what the leading order is, and so we will determine it
with the method of dominant balance.
Let the expansion of $\mathcal{G}(x)$ for large $c$ be $\mathcal{G}(x)=h_{0}(c)\mathcal{G}_{0}(x)+o(h_{0}(c))$.
On substituting the asymptotic expansions for $f(x,c)$ and $\mathcal{G}(x)$ into
(\ref{eq:phi}) we obtain
\begin{equation}
h_{0}(c)\mathcal{G}''_{0}(x)+h_{0}(c)^2\left(\mathcal{G}'_{0}(x)\right)^{2}+c^{2}f_{0}(x)+o\big(h_{0}(c)^2\big)+o(c^{2})=0
\end{equation}
We could try and cancel out the $c^{2}f_{0}(x)$ term with
$h_{0}(c)\mathcal{G}''_{0}(x)$; that would give $h_{0}=c^{2}$, but
then $h_{0}(c)\mathcal{G}''_{0}(x)$ would be subdominant to
$h_{0}^2(\mathcal{G}'_{0})^{2}$. The other option is to cancel the
$c^{2}f_{0}(x)$ term with  $h_{0}^2(\mathcal{G}'_{0})^{2}$
instead. This gives $h_{0}=c$, which  works. We therefore have
that
\begin{equation}
\mathcal{G}(x)=c\mathcal{G}_{0}(x)+\mathcal{G}_{1}(x)+O(c^{-1})
\end{equation}

The resulting equation for the leading order term in $\mathcal{G}$ is
\begin{equation}
\left[\mathcal{G}'_{0}(x)\right]^{2}+f_{0}(x)=0,
\end{equation}
the solution of which is $\mathcal{G}_{0}=\pm(x-x_{0})$. The equation for
the next order in
 $c$ is
\begin{equation}
 \mathcal{G}''_{0}(x)+2\mathcal{G}'_{0}(x)\mathcal{G}'_{1}(x)+f_{1}(x)=0,
\end{equation}
 which gives
\begin{equation}
\mathcal{G}_{1}=
\mp \left[\frac{q+\indhel}{2}\log(1+x)-\frac{q-\indhel}{2}\log(1-x)\right].
\end{equation}
The physical optics approximation for the outer solution is
therefore given by
\begin{equation}\label{eq: outer solution}
\begin{aligned}
&{}_{\indhel}S_{lm\omega}^{\text{out}}(x)=(1-x)^{\alpha}(1+x)^{\beta}{}_{\indhel}y_{lm\omega}^{\text{out}}(x)
\\
&{}_{\indhel}y_{lm\omega}^{\text{out}}(x)=(1-x)^{-(2\alpha+1)/2}(1+x)^{-(2\beta+1)/2}
\times \\ &
\quad
\times
\Big[{}_{\indhel}A_{lm\omega}(1-x)^{+(q-\indhel )/2}(1+x)^{-(q+\indhel )/2}e^{+cx}+
{}_{\indhel}B_{lm\omega}(1-x)^{-(q-\indhel )/2}(1+x)^{+(q+\indhel )/2}e^{-cx}\Big]
\end{aligned}
\end{equation}
where the constant $x_{0}$ has been absorbed within ${}_{\indhel}A_{lm\omega}$ and ${}_{\indhel}B_{lm\omega}$.
This solution is valid in the region $-1+O(c^{-1})\ll x \ll +1-O(c^{-1})$.


\subsection{Matching the solutions} \label{sec:Matching the
solutions} We have found three different solutions. One of the two inner
solutions is valid in the region $-1 \leq x \lesssim
-1+O(c^{\delta})$ for any $\delta$ such that $-1\leq\delta <0$,
and the other one for $+1-O(c^{\delta}) \lesssim x \leq +1$. The
outer solution is valid for $-1+O(c^{-1})\ll x \ll +1-O(c^{-1})$.
Clearly all three solutions together span the whole region $-1\leq
x \leq +1$. There are also two regions of overlap, one close to -1
and one close to +1, where both the outer solution and one of the
inner solutions are valid. We can proceed to match the solutions
in these regions and we will do so only to leading order in $c$ as
matching to lower orders would not bring any more insight into the
behaviour of the SWSH. When the matching is completed to leading
order, the two overlap regions are given one by
$O(c^{-1})\ll 1+x \lesssim O(c^{\delta})$ and the other one by $O(c^{-1})\ll 1-x \lesssim O(c^{\delta})$.
For the overlap regions to exist it is therefore required that we choose a $\delta$ satisfying $-1<\delta <0$.

In order to obtain an expression for the inner solution in the overlap region, we expand the inner solution for $u,u^{*}\sim \infty$.
For that, we need to know how
the confluent hypergeometric functions behave when the independent variable is large. From ~\cite{bk:AS} we have
\begin{equation}
{}_1F_{1}(b,c,z) \rightarrow \frac{\Gamma(c)e^{+i\pi b}z^{-b}}{\Gamma(c-b)}+\frac{\Gamma(c)e^{z}z^{b-c}}{\Gamma(b)}, \qquad (|z|\rightarrow +\infty)
\end{equation}
when $z=|z |e^{i\vartheta}$ with $-\pi/2 < \vartheta < 3\pi/2$, which includes the case we are considering: $\vartheta=0$.
This means that the inner solution valid close to $x=+1$ behaves like
\begin{equation} \label{eq: inner +1 asympt}
\begin{aligned}
{}_{\indhel}y_{lm\omega}^{\text{inn},+1}\rightarrow
{}_{\indhel}C_{lm\omega}
\left \{
\begin{array}{ll}
\displaystyle\frac{\Gamma(|m+\indhel |+1)\left[2c(1-x)\right]^{(-p-|m+\indhel  |-1)}e^{+c(1-x)}}{\Gamma(-p)},        & p\notin \mathbb{Z}^{+}\cup\{0\} \\
\displaystyle\frac{\Gamma(|m+\indhel |+1)e^{-i\pi p}\left[2c(1-x)\right]^{p}e^{-c(1-x)}}{\Gamma(|m+\indhel  |+1+p)},  & p\in
\mathbb{Z}^{+}\cup\{0\}
\end{array}
\right \}
\\
\qquad \qquad \qquad \qquad   , (|u|\rightarrow+\infty)
\end{aligned}
\end{equation}
The behaviour of the inner solution valid close to $x=-1$ is
similarly obtained by simultaneously replacing $x$ with $-x$,
$\indhel $ with $-\indhel $ (which also implies replacing $p$ by $p'$) 
and ${}_{\indhel}C_{lm\omega}$ with ${}_{\indhel}D_{lm\omega}$ above.

On the other hand, in order to obtain an expression for ${}_{\indhel}y_{lm\omega}^{\text{out}}$ valid in the overlap region
we perform a Taylor series expansion around $x=+1$ or $-1$ depending on where we are
doing the matching, and keep only the first order in the series:

\begin{itemize}

\item[a)] \textbf{Around $x=+1$}.

To first order in $(1-x)$:
\begin{equation} \label{eq: outer +1 asympt}
\begin{aligned}
{}_{\indhel}y_{lm\omega}^{\text{out}}(x)&\sim{}_{\indhel}A_{lm\omega}(1-x)^{\left[+(q-\indhel -1)/2-\alpha\right]}2^{\left[-(q+\indhel+1)/2-\beta\right]}e^{+cx}+ \\
&+{}_{\indhel}B_{lm\omega}(1-x)^{\left[-(q-\indhel+1)/2-\alpha\right]}2^{\left[+(q+\indhel -1)/2-\beta\right]}e^{-cx} \qquad (x\rightarrow +1)
\end{aligned}
\end{equation}
By matching the inner and outer solution in the overlap region $O(c^{-1})\ll 1-x \lesssim O(c^{\delta})$, i.e.,
by matching equations (\ref{eq: inner +1 asympt}) and (\ref{eq: outer +1 asympt}), we obtain the following relations
depending on the value of $p$:

\begin{itemize}
\item[a1)] if $p\notin \mathbb{Z}^{+}\cup\{0\}$:
\begin{equation} \label{eq: match a1}
\begin{cases}
{}_{\indhel}A_{lm\omega}=0  \\
\displaystyle
{}_{\indhel}B_{lm\omega}=2^{\left[-(q+\indhel -1)/2+\beta\right]}\frac{\Gamma(|m+\indhel |+1)}{\Gamma(-p)}(2c)^{\left[-p-|m+\indhel |-1\right]}e^{+c}{}_{\indhel}C_{lm\omega}
\end{cases}
\end{equation}

\item[a2)] if $p\in \mathbb{Z}^{+}\cup\{0\}$:
\begin{equation} \label{eq: match a2}
{}_{\indhel}A_{lm\omega}=2^{\left[+(q+\indhel+1)/2+\beta\right]}\frac{\Gamma(|m+\indhel |+1)}{\Gamma(|m+\indhel |+1+p)}e^{-i\pi p}(2c)^{p}e^{-c}{}_{\indhel}C_{lm\omega}
\end{equation}
\end{itemize}

\item[b)] \textbf{Around $x=-1$} (similar to the $x=+1$ case).

To first order in $(1+x)$:
\begin{equation}
\begin{aligned}
{}_{\indhel}y_{lm\omega}^{\text{out}}(x)&\sim{}_{\indhel}A_{lm\omega}(1+x)^{\left[-(q+\indhel+1)/2-\beta\right]}2^{\left[+(q-\indhel -1)/2-\alpha\right]}e^{+cx}+   \\
&+{}_{\indhel}B_{lm\omega}(1+x)^{\left[+(q+\indhel -1)/2-\beta\right]}2^{\left[-(q-\indhel+1)/2-\alpha\right]}e^{-cx} \qquad (x\rightarrow -1)
\end{aligned}
\end{equation}

\begin{itemize}
\item[b1)] if $p'\notin \mathbb{Z}^{+}\cup\{0\}$:
\begin{equation} \label{eq: match b1}
\begin{cases}
{}_{\indhel}B_{lm\omega}=0  \\
\displaystyle
{}_{\indhel}A_{lm\omega}=
2^{\left[-(q-\indhel -1)/2+\alpha\right]}\frac{\Gamma(|m-\indhel |+1)}{\Gamma(-p')}(2c)^{\left[-p'-|m-\indhel |-1\right]}e^{+c}{}_{\indhel}D_{lm\omega}
\end{cases}
\end{equation}

\item[b2)] if $p'\in \mathbb{Z}^{+}\cup\{0\}$:
\begin{equation} \label{eq: match b2}
{}_{\indhel}B_{lm\omega}=2^{\left[+(q-\indhel+1)/2+\alpha\right]}\frac{\Gamma(|m-\indhel |+1)}{\Gamma(|m-\indhel |+1+p')}e^{-i\pi p'}(2c)^{p'}e^{-c}{}_{\indhel}D_{lm\omega}
\end{equation}
\end{itemize}

\end{itemize}

\begin{figure}
\begin{center}
\setlength{\unitlength}{1pt}
\begin{picture}(400,150)
\put(0,80){\line(1,0){400}}
\put(0,70){\line(0,1){20}}
\put(0,60){\rotatebox{0}{\makebox(0,0)[c]{\small$-1$}}}
\put(30,70){\line(0,1){20}}
\put(30,60){\rotatebox{45}{\makebox(0,0)[r]{\small$-1+O(c^{-1})$}}}
\put(70,80){\makebox(0,0)[l]{\colorbox{red}{\makebox[44pt][c]{}}}}
\put(70,60){\rotatebox{45}{\makebox(0,0)[r]{\small$-1+O(c^{\epsilon})$}}}
\put(70,70){\line(0,1){20}}
\put(120,70){\line(0,1){20}}
\put(120,60){\rotatebox{45}{\makebox(0,0)[r]{\small$-1+O(c^{\delta})$}}}
\put(200,70){\line(0,1){20}}
\put(200,60){\rotatebox{0}{\makebox(0,0)[c]{\small$0$}}}
\put(280,80){\makebox(0,0)[l]{\colorbox{red}{\makebox[44pt][c]{}}}}
\put(280,60){\rotatebox{45}{\makebox(0,0)[r]{\small$+1-O(c^{\delta})$}}}
\put(280,70){\line(0,1){20}}
\put(330,70){\line(0,1){20}}
\put(330,60){\rotatebox{45}{\makebox(0,0)[r]{\small$+1-O(c^{\epsilon})$}}}
\put(370,70){\line(0,1){20}}
\put(370,60){\rotatebox{45}{\makebox(0,0)[r]{\small$+1-O(c^{-1})$}}}
\put(400,70){\line(0,1){20}}
\put(400,60){\rotatebox{0}{\makebox(0,0)[c]{\small$+1$}}}
\put(95,90){\makebox(0,0)[b]{\color{red}$S^{\text{match},-1}$}}
\put(305,90){\makebox(0,0)[b]{\color{red}$S^{\text{match},+1}$}}
\put(200,105){\makebox(0,0)[b]{$\overbrace{\text{\makebox[260pt]{}}}$}}
\put(200,115){\makebox(0,0)[b]{$S^{\text{out}}$}}
\put(60,120){\makebox(0,0)[b]{$\overbrace{\text{\makebox[120pt]{}}}$}}
\put(60,130){\makebox(0,0)[b]{$S^{\text{inn},-1}$}}
\put(340,120){\makebox(0,0)[b]{$\overbrace{\text{\makebox[120pt]{}}}$}}
\put(340,130){\makebox(0,0)[b]{$S^{\text{inn},+1}$}}
\end{picture}
\caption{Regions of validity in the $x$ axis of the various approximations to the SWSH for large $c$.
It must be $-1<\epsilon<\delta<0$.
For clarity, the mode labels have been dropped.
$S^{\text{match},\pm 1}$ refers to the asymptotic approximation valid in the overlap region (red) close to $x=\pm 1$.
The uniform solution is constructed as $S^{\text{unif}}=S^{\text{out}}+S^{\text{inn},+1}+S^{\text{inn},-1}-S^{\text{match},+1}-S^{\text{match},-1}$.}
\label{fig:regions of validity for asymptotic SWSH}
\end{center}
\end{figure}
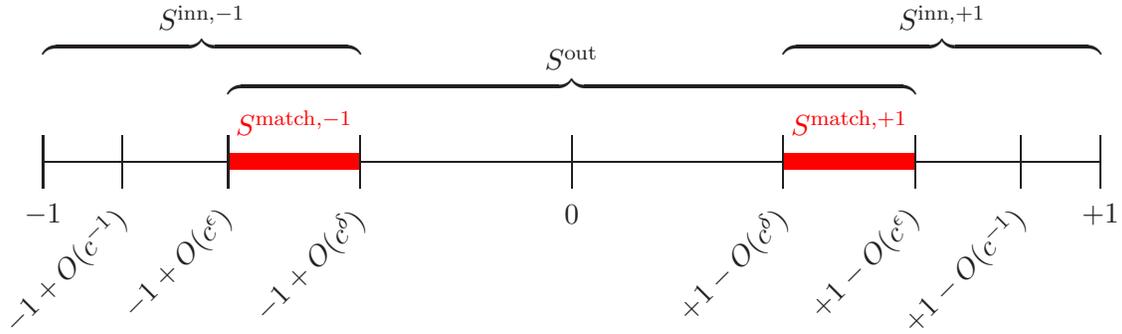

From the above matching equations we can obtain a uniform asymptotic approximation to ${}_{\indhel}S_{lm\omega}$ valid
throughout the whole region $x\in [-1,+1]$ and also find out
where the zeros of the function are. The uniform asymptotic approximation is obtained by adding the outer and the two inner solutions, and then subtracting the asymptotic
approximations in the two overlap regions since these have been included twice.
Figure \ref{fig:regions of validity for asymptotic SWSH} depicts the region of validity of the various asymptotic solutions for large $c$
that we have obtained.

We can distinguish three cases:

\subsection*{$\bm{p,p'\not\in\mathbb{Z}^{+}\cup\{0\}}$}

 From equations (\ref{eq: match a1}) and
(\ref{eq: match b1}) it must be
${}_{\indhel}A_{lm\omega}={}_{\indhel}B_{lm\omega}=0={}_{\indhel}C_{lm\omega}={}_{\indhel}D_{lm\omega}$, so this case
is the trivial solution and we discard it.

\subsection*{$\bm{p\in \mathbb{Z}^{+}\cup\{0\}}$ and $\bm{p'\not\in
\mathbb{Z}^{+}\cup\{0\}}$ , or vice-versa}

Either ${}_{\indhel}A_{lm\omega}$ or ${}_{\indhel}B_{lm\omega}$ is equal to zero (but not
both), so that the function ${}_{\indhel}S_{lm\omega}$ cannot have a zero close
to $x=0$. All the zeros, if there are any, of ${}_{\indhel}S_{lm\omega}$ are
zeros
of the inner solutions and thus they are located inside the boundary layers, close to $x=\pm 1$. \\ \\
In this case we can already directly obtain the uniform asymptotic approximation, up to an overall normalization constant ${}_{\indhel}C_{lm\omega}$:
\begin{equation} \label{eq:unif S,p,p' not}
\begin{aligned}
&{}_{\indhel}S_{lm\omega}^{\text{unif}}={}_{\indhel}C_{lm\omega}(1-x)^{\alpha}(1+x)^{\beta}
\Bigg\{e^{-c(1-x)}{}_1F_{1}\Big(-p,2\alpha+1,2c(1-x)\Big) +
\\
&+\frac{\Gamma(2\alpha+1)}{\Gamma(2\alpha+1+p)}\frac{\Gamma(-p')}{\Gamma(2\beta+1)}e^{-i\pi p}(2c)^{p+p'+2\beta+1}e^{-2c}2^{(q+\beta-\alpha)}
e^{-c(1+x)}
\times \\ &\times
{}_1F_{1}\Big(-p',2\beta+1,2c(1+x)\Big)+
2^{\left[+(q+\indhel+1)/2+\beta\right]}\frac{\Gamma(2\alpha+1)}{\Gamma(2\alpha+1+p)}e^{-i\pi p}(2c)^{p}e^{-c}e^{+cx}
\times \\ &\times
\left[(1-x)^{+(q-\indhel -1)/2-\alpha}(1+x)^{-(q+\indhel+1)/2-\beta}-2^{\left[-(q+\indhel+1)/2-\beta\right]}(1-x)^{+(q-\indhel -1)/2-\alpha}- \right.
\\
&\left. -2^{\left[+(q-\indhel -1)/2-\alpha\right]}(1+x)^{-(q+\indhel+1)/2-\beta}\right]\Bigg\}
\qquad \qquad \text{when } p\in \mathbb{Z}^{+}\cup\{0\} \text{ and } p'\notin \mathbb{Z}^{+}\cup\{0\}
\end{aligned}
\end{equation}
The uniform approximation when $p\notin \mathbb{Z}^{+}\cup\{0\}$ and $p'\in \mathbb{Z}^{+}\cup\{0\}$ may be obtained by
making the substitutions $x\leftrightarrow -x$ and $\indhel \leftrightarrow -\indhel $ (which imply the substitutions
$\alpha\leftrightarrow \beta$ and $p\leftrightarrow p'$) in (\ref{eq:unif S,p,p' not}).

The irregularity arising from
$e^{-c(1+x)}{}_1F_{1}(-p',2\beta+1,2c(1+x))\sim e^{2c}$
(ignoring factors independent of $x$ and $c$) in the limit
$x\rightarrow +1$ and $c \rightarrow +\infty$ prompted BRW to
discard the case $p'\notin \mathbb{Z}^{+}\cup\{0\}$.
It is clear from (\ref{eq:unif S,p,p' not}), however, that
this irregularity is nullified by the factor $e^{-2c}$ in front of it,
brought in by the coefficient ${}_{\indhel}D_{lm\omega}$.
Note that despite the factor $e^{-2c}$,
close to $x=-1$ this term (which is part of the inner solution
valid in the boundary layer there) is not dominated by the first
term in (\ref{eq:unif S,p,p' not}) (which is the inner solution
valid in the boundary layer near $x=+1$). The reason is that
$e^{-c(1+x)}{}_1F_{1}(-p',2\beta+1,2c(1+x))\sim e^{-2c}$ and
$e^{-c(1-x)}{}_1F_{1}(-p,2\alpha+1,2c(1-x)) \sim e^{-2c}$
where both limits are $x\rightarrow -1$ and $c \rightarrow +\infty$
and we have ignored factors independent of $x$ and $c$.
In the boundary layer around $x=\pm 1$, the asymptotic approximation valid in the overlap region close to $x=\mp 1$
cancels out the inner solution ${}_{\indhel}S_{lm\omega}^{\text{inn},\mp 1}$ in expression (\ref{eq:unif S,p,p' not}).
Similarly, in the same boundary layer,
the asymptotic approximation valid in the overlap region close to $x=\pm 1$
cancels out the outer solution, so that only ${}_{\indhel}S_{lm\omega}^{\text{inn},\pm 1}$
contributes to the uniform approximation in that boundary layer. A similar
reasoning can be applied to the case $p\notin
\mathbb{Z}^{+}\cup\{0\}$.

\subsection*{$\bm{p,p'\in \mathbb{Z}^{+}\cup\{0\}}$}
 In this case, apart
from the overall normalization constant there is another unknown
constant. We are going to determine this extra unknown by imposing
the appropriate parity under $\left\{ x\leftrightarrow -x, \indhel\leftrightarrow -\indhel  \right \}$.
Using the Teukolsky-Starobinski\u{\i} identities
(\ref{eq:Teuk-Starob. ids.s=1/2}), (\ref{eq:Teuk-Starob. ids.}) and (\ref{eq:Teuk-Starob ids. for spher.s=2})
together with the symmetry (\ref{eq:S symm.->pi-t,-s})
in the outer solution (\ref{eq: outer solution}) we obtain
\begin{equation} \label{eq: ratio D/C spin1/2}
\begin{aligned}
\frac{{}_{-\frac{1}{2}}D_{lm\omega}}{{}_{-\frac{1}{2}}C_{lm\omega}}&=\frac{{}_{+\frac{1}{2}}C_{lm\omega}}{{}_{+\frac{1}{2}}D_{lm\omega}}=
(-1)^{(l+m)}\frac{{}_{+\frac{1}{2}}C_{lm\omega}}{{}_{-\frac{1}{2}}C_{lm\omega}}= \\
& =(-1)^{(l+m)}
\begin{cases}
\displaystyle
-\frac{\sqrt{2} \sqrt{q-m}}{m+1/2}\sqrt{c}  & \qquad \text{when} \quad m\geq +\frac{1}{2}     \\
\displaystyle
-\frac{m-1/2}{\sqrt{2} \sqrt{q-m}}\frac{1}{\sqrt{c}}   & \qquad \text{when} \quad m\leq -\frac{1}{2}
\end{cases}
\end{aligned}
\end{equation}
for spin-1/2,
\begin{equation} \label{eq: ratio D/C spin1}
\begin{aligned}
\frac{{}_{-1}D_{lm\omega}}{{}_{-1}C_{lm\omega}}&=\frac{{}_{+1}C_{lm\omega}}{{}_{+1}D_{lm\omega}}=(-1)^{(l+m)}\frac{{}_{+1}C_{lm\omega}}{{}_{-1}C_{lm\omega}}= \\
& =(-1)^{(l+m)}
\begin{cases}
\displaystyle
\frac{2 \sqrt{(q-m-1)(q-m+1)}}{m(m+1)}c  & \qquad \text{when} \quad m\geq +1     \\
\displaystyle
-\frac{\sqrt{q-1}}{\sqrt{q+1}}       &  \qquad\text{when} \quad m=0           \\
\displaystyle
\frac{m(m-1)}{2 \sqrt{(q-m-1)(q-m+1)}}\frac{1}{c}  & \qquad \text{when} \quad m\leq -1
\end{cases}
\end{aligned}
\end{equation}
for spin-1 and
\begin{equation} \label{eq: ratio D/C spin2}
\begin{aligned}
&\frac{{}_{-2}D_{lm\omega}}{{}_{-2}C_{lm\omega}}=\frac{{}_{+2}C_{lm\omega}}{{}_{+2}D_{lm\omega}}=(-1)^{(l+m)}\frac{{}_{+2}C_{lm\omega}}{{}_{-2}C_{lm\omega}}=  
\\ &=
(-1)^{(l+m)}
\begin{cases}
\displaystyle
\frac{4 \sqrt{(q-m-1)(q-m+1)(q-m-3)(q-m+3)}}{(m+2)(m+1)m(m-1)}c^{2}  & \text{when} \quad m\geq +2     \\
\displaystyle
-\frac{\sqrt{q(q-2)(q-4)}}{3\sqrt{q+2}}c       & \text{when} \quad m=+1           \\
\displaystyle
\frac{\sqrt{(q-3)(q-1)}}{\sqrt{(q+3)(q+1)}}       & \text{when} \quad m=0           \\
\displaystyle
-\frac{3 \sqrt{q-2}}{\sqrt{q(q+2)(q+4)}}\frac{1}{c}       & \text{when} \quad m=-1           \\
\displaystyle
\frac{(m+1)m(m-1)(m-2)}{4 \sqrt{(q-m-1)(q-m+1)(q-m-3)(q-m+3)}}\frac{1}{c^{2}}  & \text{when} \quad m\geq -2
\end{cases}
\end{aligned}
\end{equation}
for spin-2.
Equations (\ref{eq: ratio D/C spin1/2})--(\ref{eq: ratio D/C spin2}) have been obtained without imposing
any restrictions on the values of $p$ or $p'$ and might therefore seem to
contradict the result from (\ref{eq: match a2}) and (\ref{eq: match b1}) [or (\ref{eq: match a1}) and (\ref{eq: match b2})]  giving an exponential behaviour with $c$
for the ratio ${}_{\indhel}D_{lm\omega}/{}_{\indhel}C_{lm\omega}$ for the case $p\in \mathbb{Z}^{+}\cup\{0\}$ and $p'\notin \mathbb{Z}^{+}\cup\{0\}$ [or viceversa].
We shall see in the next section, however, that equations (\ref{eq: ratio D/C spin1/2})--(\ref{eq: ratio D/C spin2}) can only actually be applied to the case
$p,p'\in \mathbb{Z}^{+}\cup\{0\}$ so that there is no such contradiction.

We can already determine in what cases the outer solution has a zero.
Clearly, from equations (\ref{eq: match a2}), (\ref{eq: match b2}) and (\ref{eq: ratio D/C spin1/2})--(\ref{eq: ratio D/C spin2}),
the ratio between the coefficients ${}_{\indhel}A_{lm\omega}$ and ${}_{\indhel}B_{lm\omega}$ is proportional
to a power of $c$,
where the constant of proportionality does not depend on $c$.
It then follows from the form (\ref{eq: outer solution}) of the outer solution that one exponential term will
dominate for positive $x$ and the other exponential term will dominate for negative $x$, when $c\rightarrow \infty$.
Therefore the outer solution does not possess a zero far from $x=0$ for large $c$.
The outer solution has a zero if ${}_{\indhel}A_{lm\omega}$
and ${}_{\indhel}B_{lm\omega}$ have
different sign and it does not have a zero otherwise. From equations (\ref{eq:def pp'}), (\ref{eq: match a2}),
(\ref{eq: match b2}) and (\ref{eq: ratio D/C spin1/2})--(\ref{eq: ratio D/C spin2}) we have:
\begin{equation}
sign\left(\frac{{}_{\indhel}A_{lm\omega}}{{}_{\indhel}B_{lm\omega}}\right)=
(-1)^{(p-p')}*sign\left(\frac{{}_{\indhel}C_{lm\omega}}{{}_{\indhel}D_{lm\omega}}\right)=(-1)^{(l+m)}
\end{equation}

Furthermore, we can calculate what the location of the zero of the outer solution is to leading order in $c$: by setting the outer solution
(\ref{eq: outer solution}) equal to zero and using (\ref{eq: match a2}) and (\ref{eq: match b2}) (since we have already seen that if $p$ and/or $p'$
$\notin \mathbb{Z}^{+}\cup\{0\}$ the outer solution does not have a zero) we obtain that for large frequency the zero is located at the following
value of $x$:
\begin{equation}
\begin{aligned}
x_0&=\frac{1}{2c}\log\left(-\frac{{}_{\indhel}B_{lm\omega}}{{}_{\indhel}A_{lm\omega}}\right)= \\
& =\frac{1}{2c}\log\left(-
2^{(-\indhel+\alpha-\beta)}\frac{\Gamma(|m-\indhel |+1)\Gamma(|m+\indhel |+1+p)}{\Gamma(|m+\indhel |+1)\Gamma(|m-\indhel |+1+p')}
e^{-i\pi(p'-p)}(2c)^{(p'-p)}\frac{{}_{\indhel}D_{lm\omega}}{{}_{\indhel}C_{lm\omega}}\right)
\end{aligned}
\end{equation}
Clearly, there is one zero in the region between the two boundary layers tending to the location $x=0$ as $c$ becomes large if
${}_{\indhel}A_{lm\omega}$ and ${}_{\indhel}B_{lm\omega}$ have different sign and there is not a zero if they have the same sign.

Finally, the uniform asymptotic approximation for this case is:
\begin{equation} \label{eq:unif S,p,p'}
\begin{aligned}
&{}_{\indhel}S_{lm\omega}^{\text{unif}}={}_{\indhel}C_{lm\omega}(1-x)^{\alpha}(1+x)^{\beta}\Bigg\{e^{-c(1-x)}{}_1F_{1}\Big(-p,2\alpha+1,2c(1-x)\Big)+
\\ &
+\frac{{}_{\indhel}D_{lm\omega}}{{}_{\indhel}C_{lm\omega}}e^{-c(1+x)}{}_1F_{1}\Big(-p',2\beta+1,2c(1+x)\Big)+
\\ & +
2^{\left[(q+\indhel+1)/2+\beta\right]}\frac{\Gamma(2\alpha+1)}{\Gamma(2\alpha+1+p)}e^{-i\pi p}(2c)^{p}e^{-c}e^{+cx}
\times \\ &\times
\left[(1-x)^{+(q-\indhel -1)/2-\alpha}(1+x)^{-(q+\indhel+1)/2-\beta}-2^{-\left[(q+\indhel+1)/2+\beta\right]}(1-x)^{+(q-\indhel -1)/2-\alpha}\right]+
\\ &
+\frac{{}_{\indhel}D_{lm\omega}}{{}_{\indhel}C_{lm\omega}}2^{\left[(q-\indhel+1)/2+\alpha\right]}
\frac{\Gamma(2\beta+1)}{\Gamma(2\beta+1+p')}e^{-i\pi p'}(2c)^{p'}e^{-c}e^{-cx}
\\ &
\left[(1+x)^{+(q+\indhel -1)/2-\beta}(1-x)^{-(q-\indhel+1)/2-\alpha}-2^{-\left[(q-\indhel+1)/2+\alpha\right]}(1+x)^{+(q+\indhel -1)/2-\beta}\right]\Bigg\}
\\ & \qquad \qquad \qquad \qquad \qquad \qquad \qquad \qquad \qquad \qquad \qquad\qquad \qquad \text{when } p,p'\in \mathbb{Z}^{+}\cup\{0\}
\end{aligned}
\end{equation}
where the ratio between ${}_{\indhel}D_{lm\omega}$ and ${}_{\indhel}C_{lm\omega}$ is given by
(\ref{eq: ratio D/C spin1/2})--(\ref{eq: ratio D/C spin2}).

Similar cancellations to the ones for the case $p\in
\mathbb{Z}^{+}\cup\{0\}$ and $p'\notin \mathbb{Z}^{+}\cup\{0\}$
occur in the present case for the uniform solution (\ref{eq:unif
S,p,p'}). The only difference is that now, in the boundary layer
around $x=\pm 1$, the asymptotic approximation valid in the
overlap region around $x=\mp 1$ only cancels out part of the
outer solution. The other part of the outer solution, however, is
exponentially negligible with respect to the inner solution
${}_{\indhel}S_{lm\omega}^{\text{inn},\pm 1}$.

\section{Calculation of $\gamma$} \label{sec:Evaluation of gamma}
In order to finally determine the value of $\gamma$ we only need to impose
that our asymptotic solution must have the correct number of
zeros. BRW give the number of zeros of the SWSH for non-negative
$m$ and $\indhel $. Straightforwardly generalizing their result for all possible values of
$m$ and $\indhel $ using the symmetries of the differential equation, we
have that the number of zeros of ${}_{\indhel}S_{lm\omega}$ is independent of $c$ and for $x\in (-1,1)$ is equal to
\begin{equation} \label{eq:zeros of SWSH}
\left\{
\begin{array}{ll}
l-|m| & \text{for} \quad |m|\geq |\indhel|\\
l-|\indhel| & \text{for} \quad |m|< |\indhel|
\end{array}
\right.
\end{equation}

The number of zeros of the confluent hypergeometric function is also needed, and that is given by Buchholz ~\cite{bk:Buchholz}:

The number of positive, real zeros of ${}_1F_{1}(-a,b,z)$ when $b>0$ is
\begin{equation}
\left\{
\begin{array}{ll}
-[-a] & \text{for} \quad  +\infty >a \geq 0 \\
0     & \text{for} \quad  0 \geq a >-\infty
\end{array}
\right.
\end{equation}
where $[n]$ means the largest integer $\leq n$.

Since the confluent hypergeometric functions are part of the inner
solutions and the region of validity of these solutions becomes
tighter to the boundary points as $c$ increases, the zeros of
${}_1F_{1}(-p,2\alpha+1,u)$ are grouped together close
to $x=+1$, and likewise for ${}_1F_{1}(-p',2\beta+1,u^{*})$ close
to $x=-1$. Apart from these zeros, for
large $c$ the function ${}_{\indhel}S_{lm\omega}$ may only have other zeros at $x=\pm 1$ and/or at
$x=x_0$. The possible one at $x=x_0$ is not due to the confluent
hypergeometric functions but to the outer solution. We define the
variable $z_{0}$ so that it has value $+1$ if ${}_{\indhel}S_{lm\omega}$ has a
zero at $x=x_0$ and value $0$ if it does not.


From equation (\ref{eq:def pp'}) we see that $p'=p+(|m+\indhel |+2\indhel -|m-\indhel |)/2$,
and therefore if either $p$ or $p'$ is integer then the other one must be integer as
well. But, as we saw in Section \ref{sec:Matching the solutions}, at least one of $p$ and $p'$
(if not both) must be a positive integer or zero.
Therefore both $p$ and $p'$ must be integers and at least one of them is positive or zero. It also follows from (\ref{eq:def pp'}) that
\begin{equation}
\gamma=2(p+p')+2+|m+\indhel |+|m-\indhel |=2q
\end{equation}
where it is now clear that $q\in \mathbb{Z}$.

Requiring that the number of zeros of the asymptotic solution coincides with the number of zeros of the
SWSH results in the condition

\begin{equation} \label{eq:equal number of zeros}
\begin{aligned}
&\left\{
\begin{array}{ll}
-(|m+\indhel |+\indhel+1)/2+q/2 & \text{for} \quad q \geq |m+\indhel |+\indhel+1 \\
0                    & \text{for} \quad q < |m+\indhel |+\indhel+1
\end{array}
\right\}+ \\
+&\left\{
\begin{array}{ll}
-(|m-\indhel |-\indhel+1)/2+q/2 & \text{for} \quad q \geq |m-\indhel |-\indhel+1 \\
0                  & \text{for} \quad q < |m-\indhel |-\indhel+1
\end{array}
\right\}+ \\
+&
z_{0}= \left\{
\begin{array}{ll}
l-|m| & \text{for} \quad |m|\geq |\indhel|\\
l-|\indhel| & \text{for} \quad |m|< |\indhel|
\end{array}
\right\}
\end{aligned}
\end{equation}

From (\ref{eq:equal number of zeros}) and the fact that $z_{0}=0$ when either $p$ or $p'\notin \mathbb{Z}^{+}\cup\{0\}$ as seen in
Section \ref{sec:Matching the solutions}, we obtain the value of $q$ in all different cases:

\begin{subequations}  \label{eq:val. of q}
\begin{align}
\begin{split}
q&=\left\{
\begin{array}{ll}
l-|m| &\text{for} \quad |m|\geq |\indhel|\\
l-|\indhel| &\text{for} \quad |m|<|\indhel|
\end{array}
\right\}
+\frac{(|m+\indhel |+|m-\indhel |)}{2}+1-z_{0}
\quad\\
&\hspace{7cm} \text{if} \quad l \geq l_{1},l_{2} \ (\text{i.e., } p,p'\in \mathbb{Z}^{+}\cup\{0\}) \label{eq:1st q} \\
\end{split} \\ \begin{split}
q&=2\left\{
\begin{array}{ll}
l-|m| &\text{for} \quad |m|\geq |\indhel|\\
l-|\indhel| &\text{for} \quad |m|<|\indhel|
\end{array}
\right\}
+|m+\indhel |+\indhel+1
\quad \\
&\hspace{7cm}\text{if} \quad l<l_{2} \ (\text{i.e., } p\in,p'\notin \mathbb{Z}^{+}\cup\{0\})  \label{eq:2nd q} \\
\end{split} \\ \begin{split}
q&=2\left\{
\begin{array}{ll}
l-|m| &\text{for} \quad |m|\geq |\indhel|\\
l-|\indhel| &\text{for} \quad |m|<|\indhel|
\end{array}
\right\}
+|m-\indhel |-\indhel+1
\quad \\
&\hspace{7cm}\text{if} \quad l<l_{1} \ (\text{i.e., } p\notin,p'\in
\mathbb{Z}^{+}\cup\{0\})  \label{eq:3rd q}
\end{split}
\end{align}
\end{subequations}

where \\ \\
$l_{1}\equiv \left\{ \begin{array}{ll} |m| &\text{for} \quad |m|\geq |\indhel|\\ |\indhel| &\text{for} \quad |m|<|\indhel|\end{array} \right\}+
(|m+\indhel |-|m-\indhel |)/2+\indhel $ \\
$l_{2}\equiv \left\{ \begin{array}{ll} |m| &\text{for} \quad |m|\geq |\indhel|\\ |\indhel| &\text{for} \quad |m|<|\indhel| \end{array} \right\}+
(|m-\indhel |-|m+\indhel |)/2-\indhel $ \\

By requiring in (\ref{eq:1st q}) that $q$ must also satisfy (\ref{eq:def pp'}) and bearing in mind that $z_{0}$ can only have the values $0$ or $1$, it must
be
\begin{equation} \label{eq:z0=0,1 if S has zero at x=0 or not}
z_{0}=
\begin{cases}
0 & \text{for} \quad l-l_{1} \quad \text{even}   \\
1 & \text{for} \quad l-l_{1} \quad \text{odd}
\end{cases}
\end{equation}
where $l_{2}$ instead of $l_{1}$ could have been used, since one
is equal to the other one plus an even number.


It can be trivially seen that if $l_{1}$ has an allowed value,
i.e.,
\begin{equation}
l_{1}\geq \left\{ \begin{array}{ll} |m| &\text{for} \quad |m|\geq
|\indhel|\\ |\indhel| &\text{for} \quad |m|<|\indhel|\end{array} \right\},
\end{equation}
then $l_{2}$ does not, and vice-versa, so that cases (\ref{eq:2nd
q}) and (\ref{eq:3rd q}) are mutually exclusive.

Clearly, when $l<l_{1}$ or $l<l_{2}$, for fixed $\indhel $ and $m$, as $l$ is increased by $1$ the corresponding value of $q$ is also increased by $1$, so that two
different values of $l$ correspond to two different values of $q$. However, once the threshold $l \geq \max(l_{1},l_{2})$ is reached, every increase of $2$ in $l$
will involve the subtraction of an extra $1$ in (\ref{eq:1st q}) via $z_{0}$, so that its corresponding value of $q$ will be the same as for the previous $l$.
Therefore, in the region $l \geq \max(l_{1},l_{2})$, every value of $q$ will correspond to two consecutive, different $l$'s: the two corresponding SWSH's
will have the same number of zeros and behaviour close to the boundary points, but one will have a zero at $x=x_0$ and the other one will not.

Another feature that can be seen is that, for $\indhel = \pm \frac{1}{2}$, the case
$l < l_{1}$ or $l_{2}$ (i.e., $p$ or $p'\notin
\mathbb{Z}^{+}\cup\{0\}$) implies $q-m=0$ or $l<|m|$ when
$m\geq \frac{1}{2}$ and $m\leq -\frac{1}{2}$ respectively, so that (\ref{eq: ratio D/C spin1/2}) 
is not applicable to these cases, as already mentioned in the previous section.

Similarly, for $\indhel = \pm 1$, the case
$l < l_{1}$ or $l_{2}$ implies $q-m=\pm 1$ or $q=+1$ when
$m\geq 1$ and $m=0$ respectively, so that (\ref{eq: ratio D/C
spin1}) is not valid for these cases. When $m\leq -1$ it follows from (\ref{eq:2nd q})
and (\ref{eq:3rd q}) that $l < l_{1}$ or $l_{2}$ requires $l<|m|$,
which is not allowed.

For $\indhel = \pm 2$, $l < l_{1}$ or $l_{2}$ implies
\begin{equation*}
\begin{cases}
m-q=\pm 1,\pm 3& \text{when} \quad m \geq 2   \\
q=0,2,4 & \text{when} \quad m=1 \\
q=1,3& \text{when} \quad m=0 \\
q=2& \text{when} \quad m=-1 \\
l<|m|& \text{when} \quad m\leq-2 \\
\end{cases}
\end{equation*}
so that (\ref{eq: ratio D/C spin2}) is not valid then.

Note that the scalar case is obtained from our formulae as a
particular case. Setting $\indhel =0$ in the equations above we have
$l_{1}=l_{2}=|m|$ and therefore $l$ will always be greater or
equal than both $l_{1}$ and $l_{2}$ so that (\ref{eq:1st q}) will
apply, and it gives $q=l+1-z_{0}$ with
\begin{equation*}
z_{0}=
\left\{
\begin{array}{ll} 0 &\text{for} \quad l-|m| \quad \text{even}   \\
1 &\text{for}  \quad l-|m| \quad \text{odd} \end{array} \right\}.
\end{equation*}
We also have have $p=p'\in  \mathbb{Z}^{+}\cup\{0\}$ and $2\alpha
=2\beta =|m|$ and then the confluent hypergeometric functions are
just the generalized Laguerre polynomials:
${}_1F_{1}(-p,|m|+1,z)\propto L_{p}^{(|m|)}(z)$. Finally, because
of the existence of the $x\leftrightarrow -x$ symmetry in the
scalar case, we have that ${}_0B_{lm\omega}=\pm {}_0A_{lm\omega}$ in
(\ref{eq: outer solution}) and therefore the zero of the outer
solution, if it exists, will be located exactly at $x=0$. All
these results for the scalar case coincide with
~\cite{bk:high_transc_funcs},~\cite{bk:Flammer} and ~\cite{bk:Meixner&Schafke}.


\section{Numerical method} \label{sec:num. method; high freq. sph.}

Two different methods have been used to obtain the numerical data. One method is the one used by Sasaki and Nakamura~\cite{ar:Sasa&Naka'82}, consisting in
approximating the differential equation (\ref{eq:ang. teuk. eq. for y}) by a finite difference equation,
and then finding the eigenvalue as the value of ${}_{\indhel}E_{lm\omega}$
that makes zero the determinant of the resulting (tri-diagonal) matricial equation.
We have used this method to find the eigenvalues for several large values of $c$.
However, we used the shooting method described in ~\cite{bk:NumRec} to calculate the spin-weighted spheroidal function.

The shooting method is applied in ~\cite{bk:NumRec} to the spheroidal differential equation (i.e., $\indhel=0$), and we adapted it
to the spin-weighted spheroidal differential equation as follows.
In general, for an initial, arbitrary value ${}_{\indhel}\hat{E}_{lm\omega}$ for the eigenvalue, which is different from the
actual eigenvalue ${}_{\indhel}E_{lm\omega}$, the numerically integrated solution
is a combination of both the regular and the irregular solutions, i.e.,
\begin{equation}
{}_{\indhel}y^{\text{num}}_{lm\omega}=A({}_{\indhel}\hat{E}_{lm\omega}){}_{\indhel}y_{lm\omega}+A({}_{\indhel}\hat{E}_{lm\omega}){}_{\indhel}y^{\text{irreg}}_{lm\omega}
\end{equation}
where ${}_{\indhel}y^{\text{irreg}}_{lm\omega}$ is the irregular solution at $x=\pm 1$,
${}_{\indhel}y^{\text{num}}_{lm\omega}$ is the numerically obtained solution and
${}_{\indhel}y_{lm\omega}$ is the analytic, regular solution.
$A$ and $B$ are unknown functions of ${}_{\indhel}\hat{E}_{lm\omega}$.
We need to modify the value of ${}_{\indhel}\hat{E}_{lm\omega}$ so that only the regular term $A{}_{\indhel}y_{lm\omega}$ is retained.
In the scalar case, the boundary condition at $x_2\equiv +1-\d{x}$ may be imposed by requiring that ${}_{\indhel=0}\hat{E}_{lm\omega}$ is a zero of the function
$g({}_{\indhel=0}\hat{E}_{lm\omega})\equiv {}_{\indhel=0}y^{' \text{num}}_{lm\omega}(x_2)-{}_{\indhel=0}y'_{lm\omega}(x_2)$, 
where the analytic value ${}_{\indhel}y'_{lm\omega}(x_2)$ is known for the scalar case because
${}_{\indhel=0}y'_{lm\omega}(x)\propto {}_{\indhel=0}y'_{lm\omega}(-x)$.
The function $g({}_{\indhel=0}\hat{E}_{lm\omega})$ should tend to zero as ${}_{\indhel=0}\hat{E}_{lm\omega}$ approaches the
correct eigenvalue and should tend to infinity when it is far from it because of the behaviour of the irregular solution.
However, in general we have ${}_{\indhel}y'_{lm\omega}(x)\propto {}_{-\indhel}y'_{lm\omega}(-x)$,
relating solutions of equations with different helicity when $h\neq 0$, and therefore we do
not know the analytic value ${}_{\indhel}y'_{lm\omega}(x_2)$ for a particular value $\indhel\neq 0$ of the helicity.
We therefore decided to apply the shooting method by
finding a zero of the function
\begin{equation}
g({}_{\indhel}\hat{E}_{lm\omega})\equiv {}_{\indhel}y^{' \text{num}}_{lm\omega}(x_2)-{}_{\indhel}y^{' \text{approx}}_{lm\omega}(x_2)
\end{equation}
instead, where ${}_{\indhel}y^{' \text{approx}}_{lm\omega}(x_2)$ is not the actual analytic value,
which we do not know, but an approximation to it:
\begin{equation}
{}_{\indhel}y_{lm\omega}^{' \text{approx}}(x_2)\simeq \frac{{}_{\indhel}y^{\text{num}}_{lm\omega}(x_2)}{{}_{\indhel}y_{lm\omega}(x_2)}{}_{\indhel}y'_{lm\omega}(x_2)
\end{equation}

Sasaki and Nakamura's method, which they only develop explicitly for the case $\indhel =-2$ and $m=0$ solves the
angular differential equation (\ref{eq:ang. teuk. eq. for y}) re-written with derivatives with respect to $\theta$ rather than $x$:
\begin{equation} \label{eq:ang. teuk. eq. for y with theta-derivs.}
\begin{aligned}
& \Bigg\{
 \ddiff{}{\theta}+\frac{1}{\sin\theta}\left[2(\alpha-\beta)+2(\alpha+\beta)\cos\theta+\cos\theta\right]\diff{}{\theta}+  \\
&  +{}_{\indhel}E_{lm\omega}-(\alpha+\beta)(\alpha+\beta+1)+c^{2}\cos^{2}\theta-2\indhel c\cos\theta
\Bigg\} {}_{\indhel}y_{lm\omega}(\theta)=0
\end{aligned}
\end{equation}
This equation is approximated by a finite-difference equation. Apart from at the boundaries, the derivatives are replaced with central differences.
At the boundary points, the regularity condition (\ref{eq:asympt. S for x->+/-1}) requires that $\left. \d{{}_{\indhel}y_{lm\omega}}/\d{\theta}\right|_{x=\pm 1}=0$ and
the first order derivative (which has a factor $1/\sin\theta$ in front) is approximated by a forward/backward difference at $x=+1/-1$ respectively.
The result is that equation
(\ref{eq:ang. teuk. eq. for y with theta-derivs.}) is approximated by
\begin{equation} \label {eq:finite-diff. ang. teuk. eq. for y with theta-derivs.}
\begin{aligned}
&
\frac{{}_{\indhel}y^{i+1}_{lm\omega}-2{}_{\indhel}y^{i}_{lm\omega}+{}_{\indhel}y^{i-1}_{lm\omega}}{(\Delta\theta)^2}+\\
&+\frac{1}{\sin\theta_i}\left[2(\alpha-\beta)+2(\alpha+\beta)\cos\theta_i+\cos\theta_i\right]
\frac{{}_{\indhel}y^{i+1}_{lm\omega}-{}_{\indhel}y^{i-1}_{lm\omega}}{2\Delta\theta}+ \\
&
+\left[{}_{\indhel}E_{lm\omega}-(\alpha+\beta)(\alpha+\beta+1)+c^{2}\cos^{2}\theta_i-2\indhel c\cos\theta_i
\right]{}_{\indhel}y^{i}_{lm\omega}=0,\\
&\hspace{10cm} \text{for\, } i=2,\dots,2N \\
& 2(1+2\alpha)\frac{2{}_{\indhel}y^{i+1}_{lm\omega}-2{}_{\indhel}y^{i}_{lm\omega}}{(\Delta\theta)^2}+ \\
& +\left[{}_{\indhel}E_{lm\omega}-(\alpha+\beta)(\alpha+\beta+1)+c^{2}-2\indhel c\right]{}_{\indhel}y^{i}_{lm\omega}=0,  \text{\, for\, } i=1 \, (\theta=0) \\
& -4\beta\frac{2{}_{\indhel}y^{i-1}_{lm\omega}-2{}_{\indhel}y^{i}_{lm\omega}}{(\Delta\theta)^2}+ \\
& +\left[{}_{\indhel}E_{lm\omega}-(\alpha+\beta)(\alpha+\beta+1)+c^{2}+2\indhel c\right]{}_{\indhel}y^{i}_{lm\omega}=0, \text{\, for\, } i=2N+1 \, (\theta=\pi)
\end{aligned}
\end{equation}
where
$\theta_i=\pi(i-1)/(2N)\equiv \Delta\theta(i-1)$ and
$i=1,2,\dots,2N+1$. Equation (\ref{eq:finite-diff. ang. teuk. eq.
for y with theta-derivs.}) can be represented as the product of a
square, tridiagonal matrix $A$ of dimension $(2N+1)\times(2N+1)$
and the vector of elements ${}_{\indhel}y^{i}_{lm\omega}$ equal to
zero. In order to find the eigenvalue, Sasaki and Nakamura's
method imposes that the determinant of matrix $A$ is zero.

We found that, already with $N=100$, for most modes the values of ${}_{\indhel}E_{lm\omega}$ obtained to quadruple precision  actually provided
values of the determinant so large that were even greater than the machine's largest number.
We therefore decided to use this method
only to find eigenvalues and use the shooting method
when we wish to find both eigenvalues and spherical functions.
In fact, Sasaki and Nakamura's method without finding the spherical funcion is so much faster than
the shooting method
that the former is the preferable method to use if we wish to find eigenvalues far from any known eigenvalue (as we analytically do for $c=0$ for example).
This is why we used Sasaki and Nakamura's method to find the eigenvalues for large frequency and then used the resulting eigenvalue to find
the corresponding spherical function with
the shooting method.

We wrote a program that
implements Sasaki and Nakamura's method to find eigenvalues,
particularly adapted to the case of large frequency. It calculates
${}_{\indhel}\lambda_{lm\omega}$ rather than ${}_{\indhel}E_{lm\omega}$ since
${}_{\indhel}\lambda_{lm\omega}\sim O(c)$ for large $c$ whereas
${}_{\indhel}E_{lm\omega}\sim O(c^2)$. It starts with the known
value of ${}_{\indhel}\lambda_{l,m,\omega =0}$ (\ref{eq:eigenval. for c=0}) and
finds the eigenvalue ${}_{\indhel}\lambda_{lm\omega}$ for increasing frequency
by looking for a zero of the determinant of the matrix $A$. This
procedure is smooth no matter how large the frequency is if
$l<l_1$ or $l<l_2$. However, if $l\geq \max(l_1,l_2)$, for some large
value of the frequency, the eigenvalues for two consecutive values
of $l$ are so close (since they correspond to the same $q$ and
therefore their leading order term for large frequency is the
same) that the initial bracketing of the eigenvalue includes both
eigenvalues and therefore $\det A$ calculated with the values of
${}_{\indhel}\lambda_{lm\omega}$ at the two ends of the bracket has the same
sign. From this value of the frequency on, instead of looking for
a zero of the determinant the program just looks for the value
${}_{\indhel}\lambda_{lm\omega}$ that is an extreme of the determinant. The
reason is that this provides a point which is in between the two
actual eigenvalues and it is therefore useful both as an
approximation and as a bracket point for either of them. Instead
of using minimization/maximization routines, which are very costly
in terms of accuracy and time, in order to find an extreme of
$\det A$, the program
looks for a zero of the derivative of $\det A$, which can be calculated to be
\begin{equation}
\diff{(\det A)}{{}_{\indhel}\lambda_{lm\omega}}=\mathop{\mathrm{trace}}
\left[(\det A) A^{-1}\right]
\end{equation}
and is very easy to evaluate.
The program
we have just described
provided the graphs of ${}_{\indhel}\lambda_{lm\omega}$
as a function of $\omega $ for large frequency and there is therefore no need for it to distinguish with accuracy between the two consecutive eigenvalues.

The extreme point of the determinant found by
this program
is used by another program which uses the shooting method and Runge-Kutta integration
to bracket and determine the two close eigenvalues and their corresponding angular functions.
It initially looks for a zero of the function
$g({}_{\indhel}E_{lm\omega})$ inside a bracket of the eigenvalue.
If it finds a zero inside the bracket, it then directly implements the shooting method as described in ~\cite{bk:NumRec}.
Instead, if it does not find a zero inside the bracket it
then assumes that it is because the frequency is large enough so that there are two eigenvalues inside the bracket corresponding to two different,
consecutive $l$'s.
It then looks
for a minimum of $g({}_{\indhel}E_{lm\omega})$ (with a possible change of sign if there is a maximum instead) and uses that
minimum to find a zero to its right or to its left depending on which one corresponds to the $l$ we are interested in, according to (\ref{eq:val. of q}).
This second program
also finds the zero of the function ${}_{\indhel}S_{lm\omega}$ close to $x=0$ for large $\omega $
if it has one as indicated by (\ref{eq:z0=0,1 if S has zero at x=0 or not}), uses a smaller stepsize in $x$ close to $x=\pm 1$ to
cater for the rapid oscillations of the angular function there for large $\omega $ and makes use of equations (\ref{eq:val. of q})
and (\ref{eq:series E for large w}) to help bracket the eigenvalue.
This program
provided the graphs of ${}_{\indhel}S_{lm\omega}(\theta)$ for large frequency.

Both programs were written in Fortran90 and contain parallel algorithms that use
the Message-Passing Interface as the message-passing library.


\section{Numerical results} \label{sec:num. results; high freq. sph.}

All the numerical results and graphs in this section have been obtained setting $a=0.95$ and $M=1$.

There is an obvious numerical problem when $p,p'\in
\mathbb{Z}^{+}\cup\{0\}$. In this case, as mentioned in Section
\ref{sec:Evaluation of gamma}, the eigenvalues for two different
values of $l$ (but same $\indhel ,m$) become exponentially close as
$c$ increases (~\cite{ar:BRW}). This means that for this case we are
not able to find the functions for very large values of the
frequency. For example, in the case below for $\indhel =-1$ and $m=1$, when
$\omega =25$ the eigenvalues for $l=3$ and $l=4$ only differ in their
14th digit.

\begin{figure}[p]
\rotatebox{90}
\centering
\includegraphics*[width=80mm,angle=270]{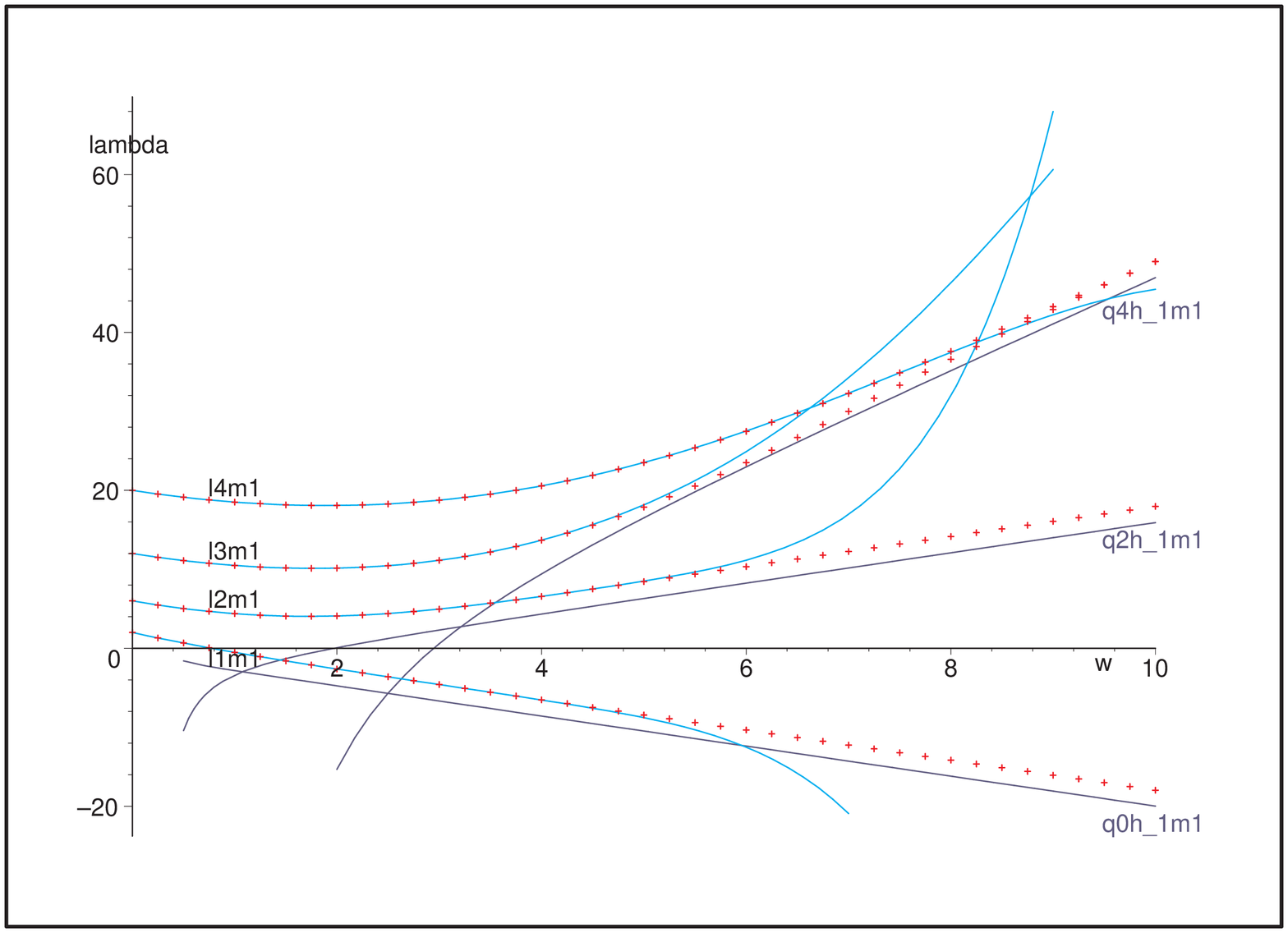}
\label{fig:lambda_s_1m1w0to10}
\\
\includegraphics*[width=80mm,angle=270]{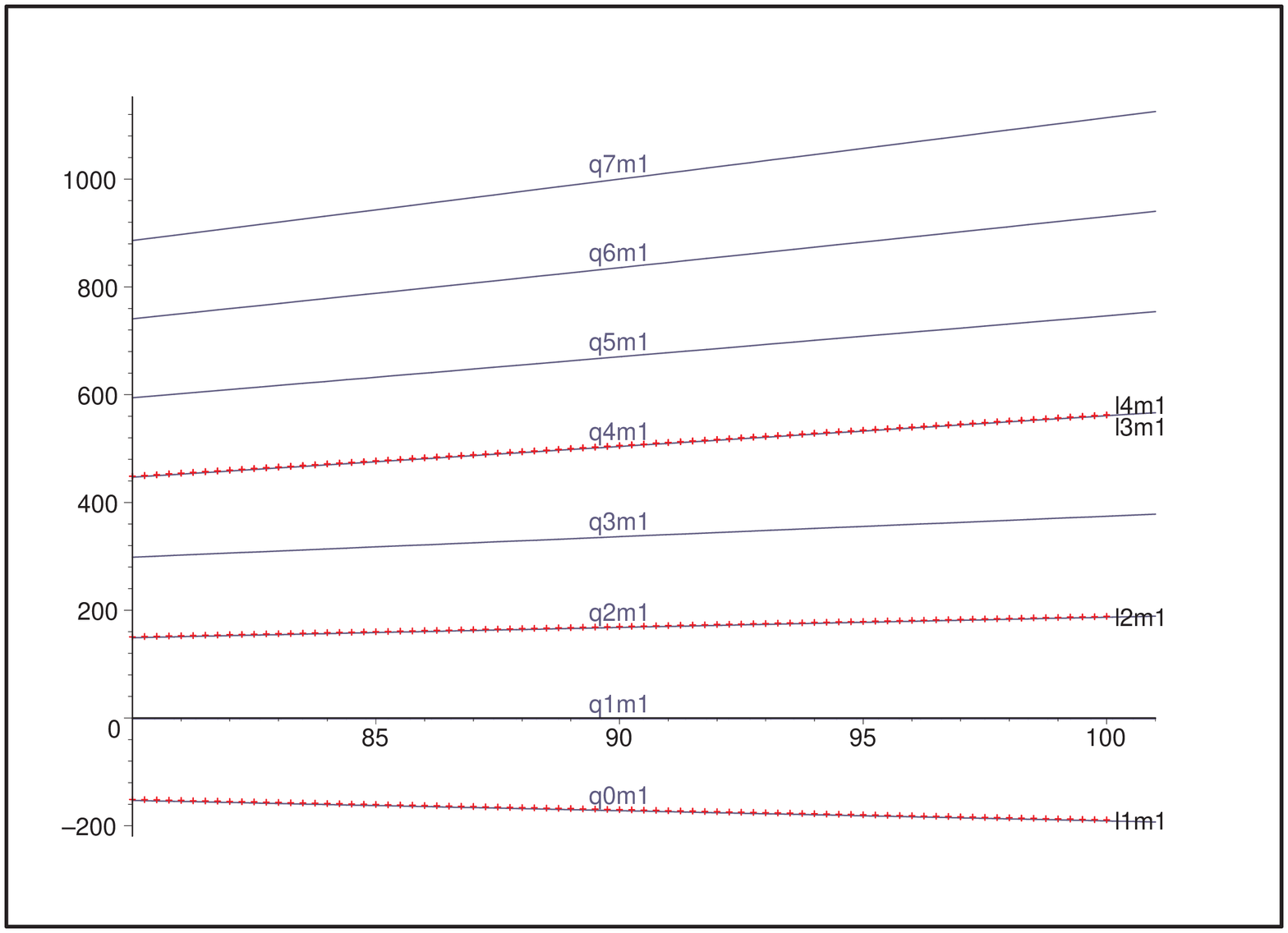}
\caption{${}_{-1}\lambda_{l,1,\omega}$ as a function of $\omega$ for several $l$ and $q$. 
The red crosses are the numerical data. 
The navy blue lines are using BRW's expansion for ${}_{\indhel}\lambda_{lm\omega}$
and the light blue lines are Press and Teukolsky's.
} \label{fig:lambda_s_1m1w0to100}
\end{figure}
\begin{figure}[p]
\rotatebox{90}
\centering
\includegraphics*[width=80mm,angle=270]{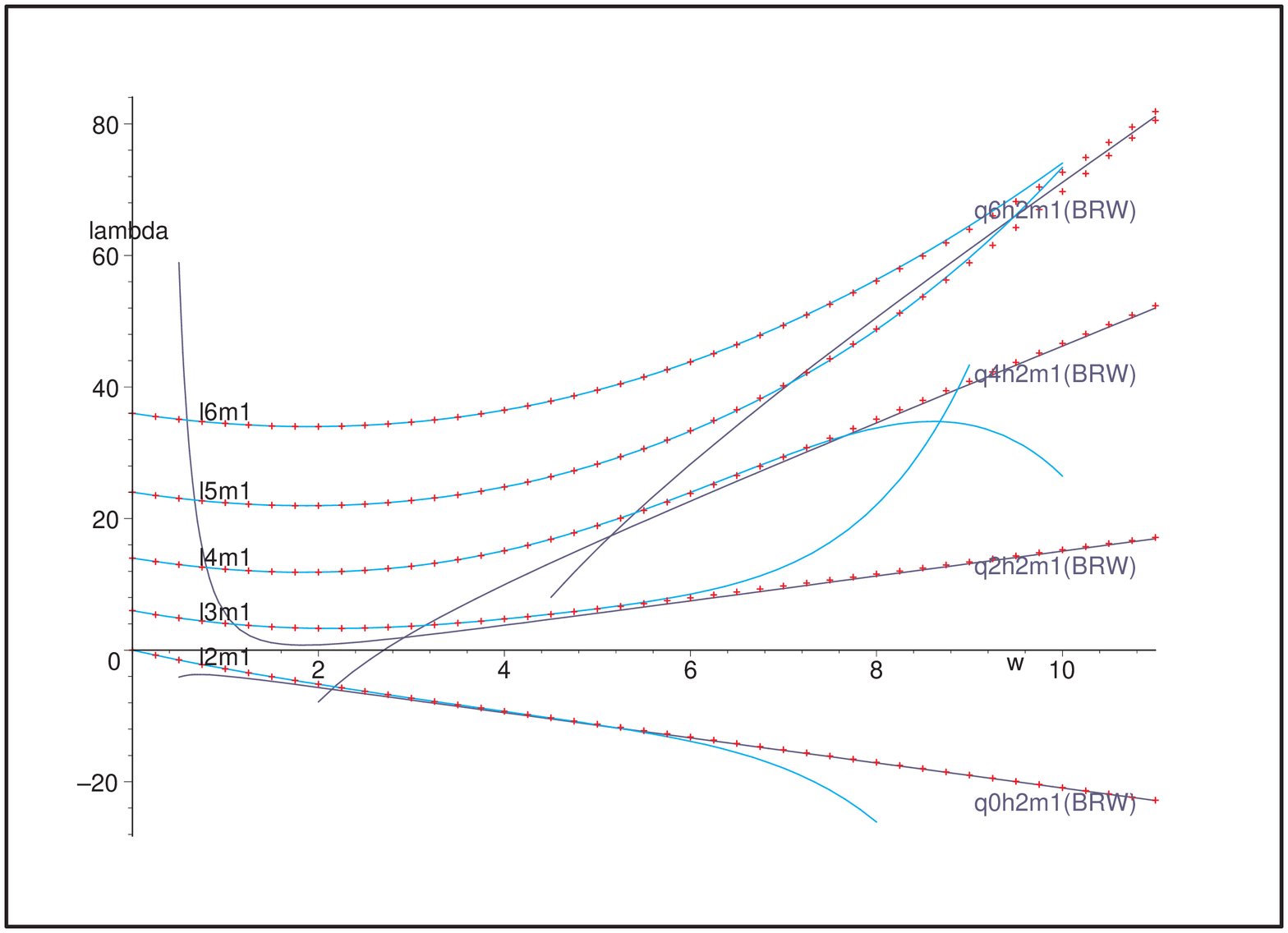}
\\
\includegraphics*[width=80mm,angle=270]{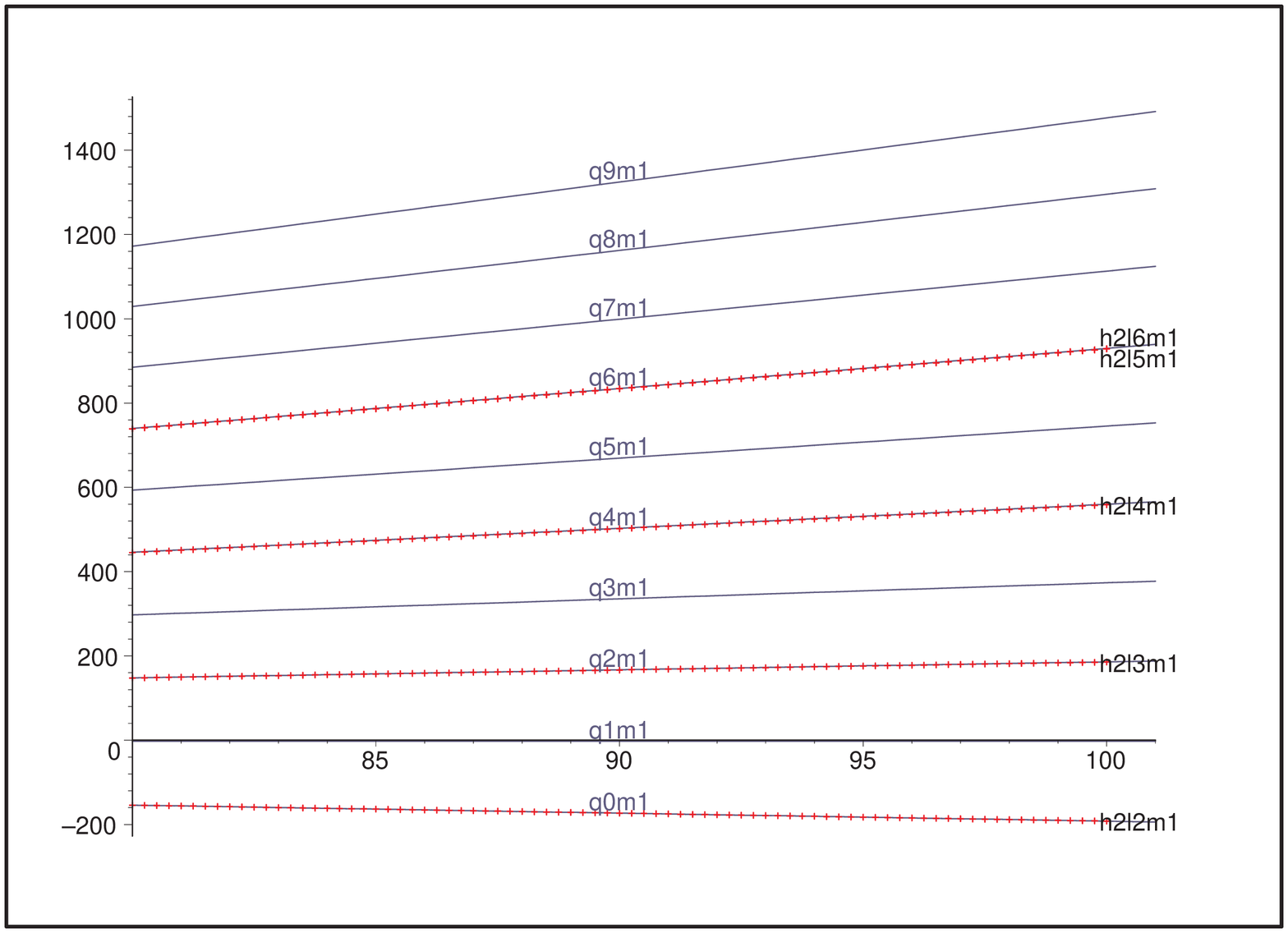}
\caption{${}_{+2}\lambda_{l,1,\omega}$ as a function of $\omega$ for several $l$ and $q$. 
The red crosses are the numerical data. 
The navy blue lines are using BRW's expansion for ${}_{\indhel}\lambda_{lm\omega}$
and the light blue lines are Press and Teukolsky's.
} \label{fig:lambda_s2m1w0to100}
\end{figure}

BRW do give the analytical value for $q$ for spin-0. For spin
different from zero, however, they try to numerically match their
large-frequency asymptotic expansion of the eigenvalue with the
expansion for small frequency given by Press and Teukolsky
(~\cite{ar:Press&Teuk'73} and ~\cite{ar:Teuk&Press'74}). As can
be seen in Figures
\ref{fig:lambda_s_1m1w0to100}
and
\ref{fig:lambda_s2m1w0to100}, this matching at intermediate
values of the frequency might be good for certain cases,
especially for
 small $l$, but not for other ones. All eigenvalues start off for frequency zero at the value given by (\ref{eq:eigenval. for c=0}), as expected, and when
$l\geq l_{1}$ or $l_{2}$ the pairs of curves that share the same
value of $q$ become exponentially closer and closer to each other
 as the frequency increases. When the frequency is as large as $100$, the curves fully coincide in
the expected pairs for large frequency (given by
equation (\ref{eq:series E for large w}), and BRW for lower order terms) where $q$ comes in as
a parameter. From this, the corresponding value of $q$ for a certain set of values of $\{l,m,\indhel\}$ can be inferred,
and this coincides with the one given by equations (\ref{eq:1st q}), (\ref{eq:2nd q}) and (\ref{eq:3rd q}).

\begin{figure}[p]
\rotatebox{90}
\centering
\includegraphics*[width=90mm,angle=270]{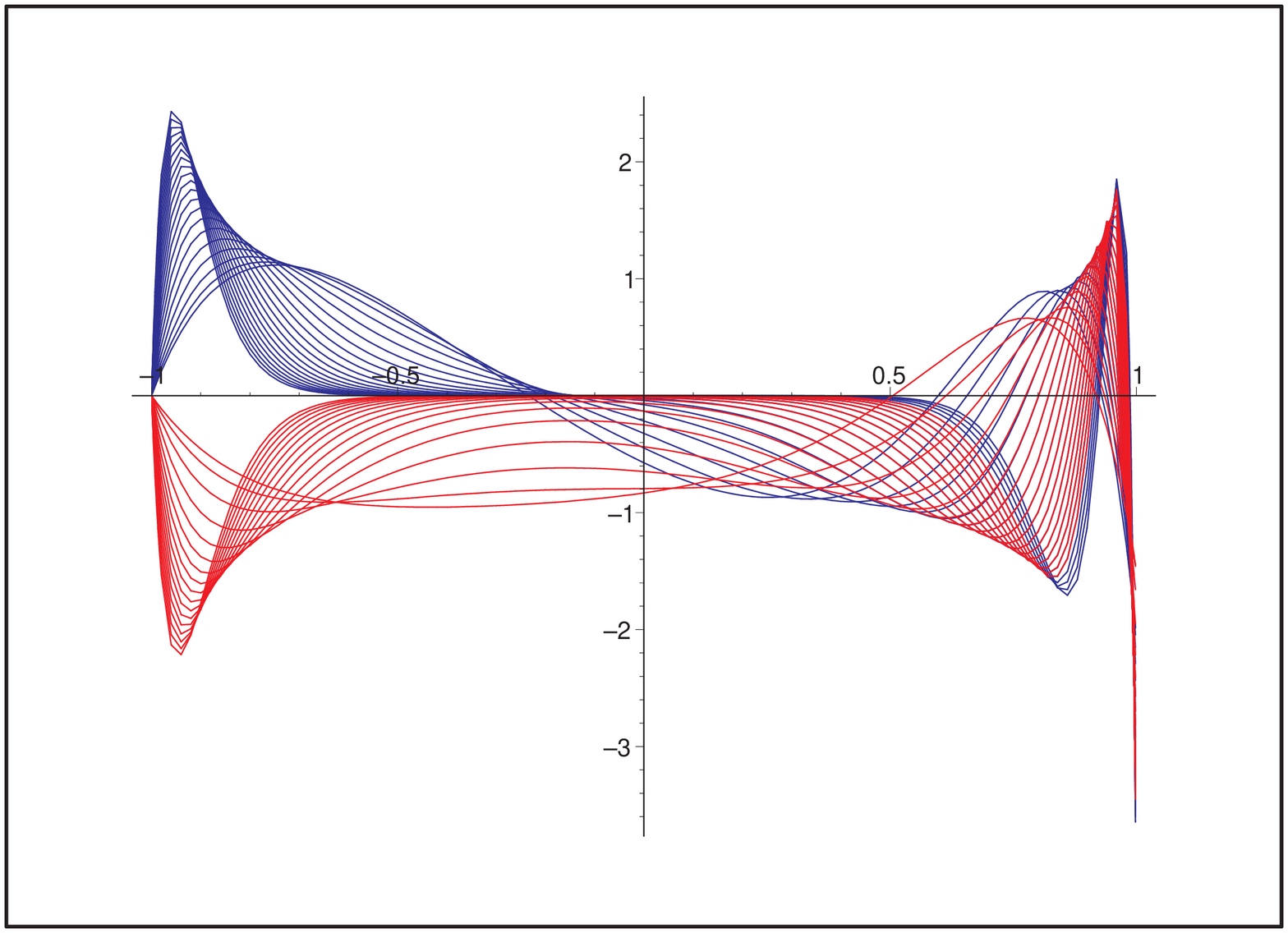}
\caption{${}_{-1}S_{l,1,\omega}$ for $l=3\ \&\ 4$, $\omega =5\rightarrow25$.
Blue lines correspond to $l=4$ and the red ones to $l=3$.
As $\omega$ increases the curves become increasingly flattened out in the region close to the origin.} \label{fig:sph_n3n4m1w5to25 first}
\includegraphics*[width=90mm,angle=270]{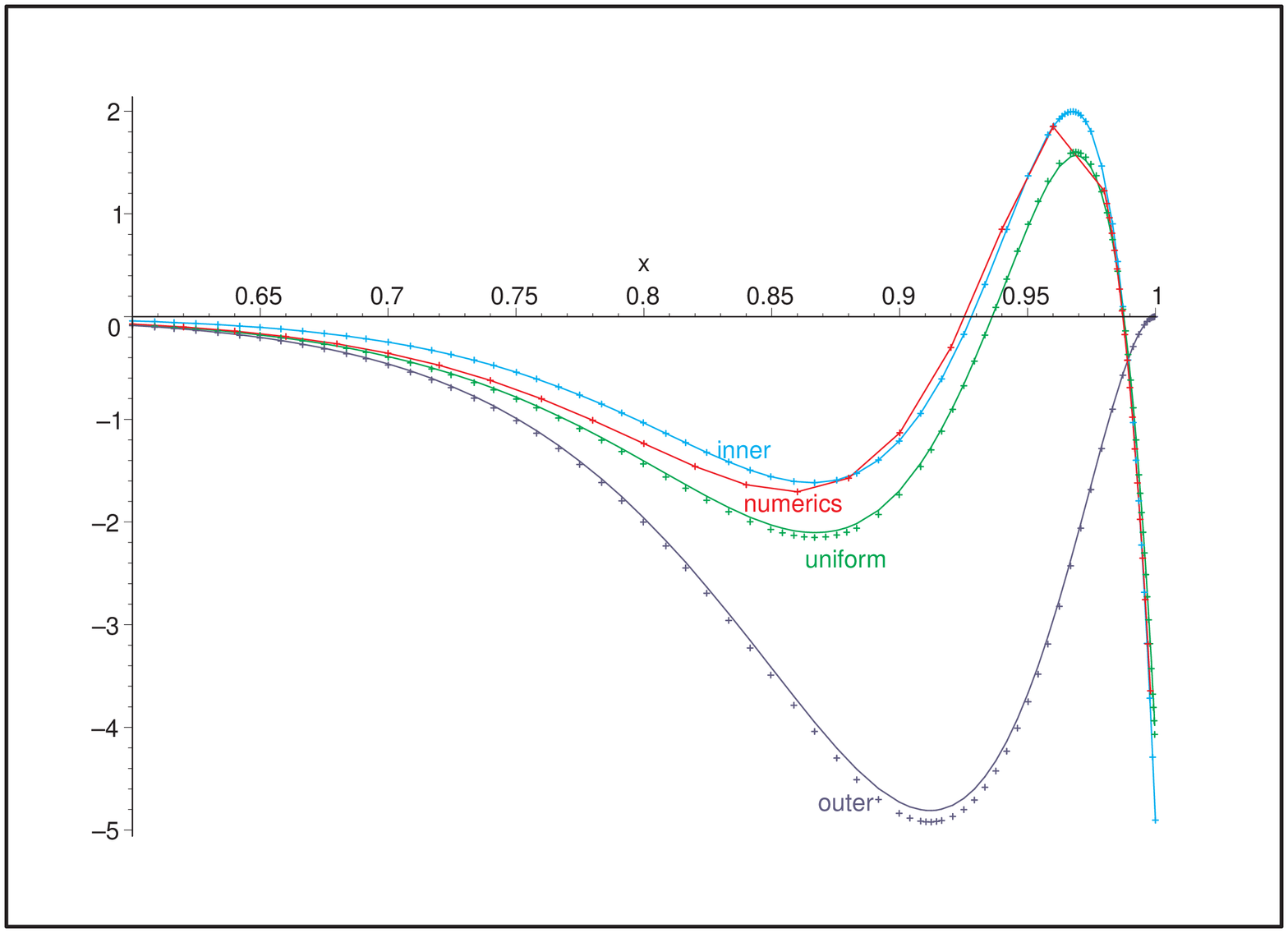}
\caption{${}_{-1}S_{l,1,25}$ for $l=3\ \&\ 4$.
Different solutions as labeled.
The continuous lines correspond to $l=4$ and the dotted ones to $l=3$.} \label{fig:sph_n3n4m1w5to25}
\end{figure}

We calculated and plotted in Figure \ref{fig:sph_n3n4m1w5to25 first} the
SWSH for $\indhel =-1$, $l=3\ \&\ 4$, $m=1$, where the value of $q$,
given by (\ref{eq:2nd q}), is the same for both of them: $q=4$
(this is a case where $p,p'\in \mathbb{Z}^{+}\cup\{0\}$).
Several features can be seen.
Firstly, as the frequency increases from $\omega =5$ to 25,
 the functions become flattened out in the middle region of $x$ and squeezed out towards the edges. Since the value of $q$ is the same for both cases, the inner solution is
the same for both of them, with the only exception of the relative sign between the inner solution for positive $x$ and negative $x$
(\ref{eq: ratio D/C spin1}). The function for $l=4$ has three zeros and the one for $l=3$ has two, in agreement with (\ref{eq:zeros of SWSH}). The inner solution provides
for the two zeros of $l=3$ and the corresponding two of $l=4$, and these become closer to the boundary point $x=+1$ as the frequency increases. The extra
zero of $l=4$ comes from the outer solution and becomes closer to $x=0$ with increasing frequency.

\begin{figure}[p]
\rotatebox{90}
\centering
\includegraphics*[width=90mm,angle=270]{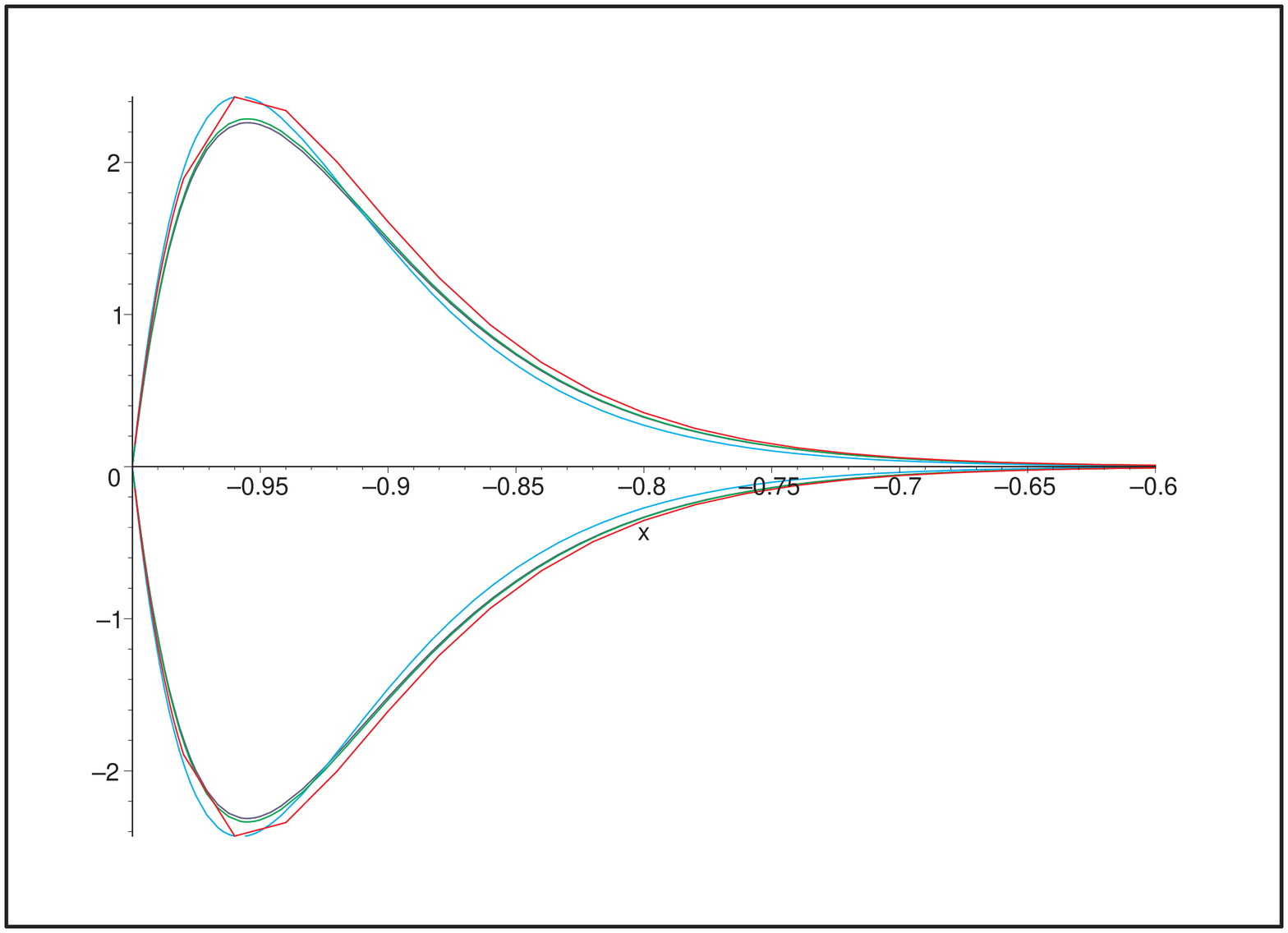}
\caption{${}_{-1}S_{l,1,25}$ for $l=3\ \&\ 4$.
The curves above the $x$-axis correspond to $l=4$ and below the axis to $l=3$.
Correspondence between colours and solutions is the same as in Figure \ref{fig:sph_n3n4m1w5to25}.} \label{fig:sph_n3n4m1w5to25_w25_x_06to_1}
\includegraphics*[width=90mm,angle=270]{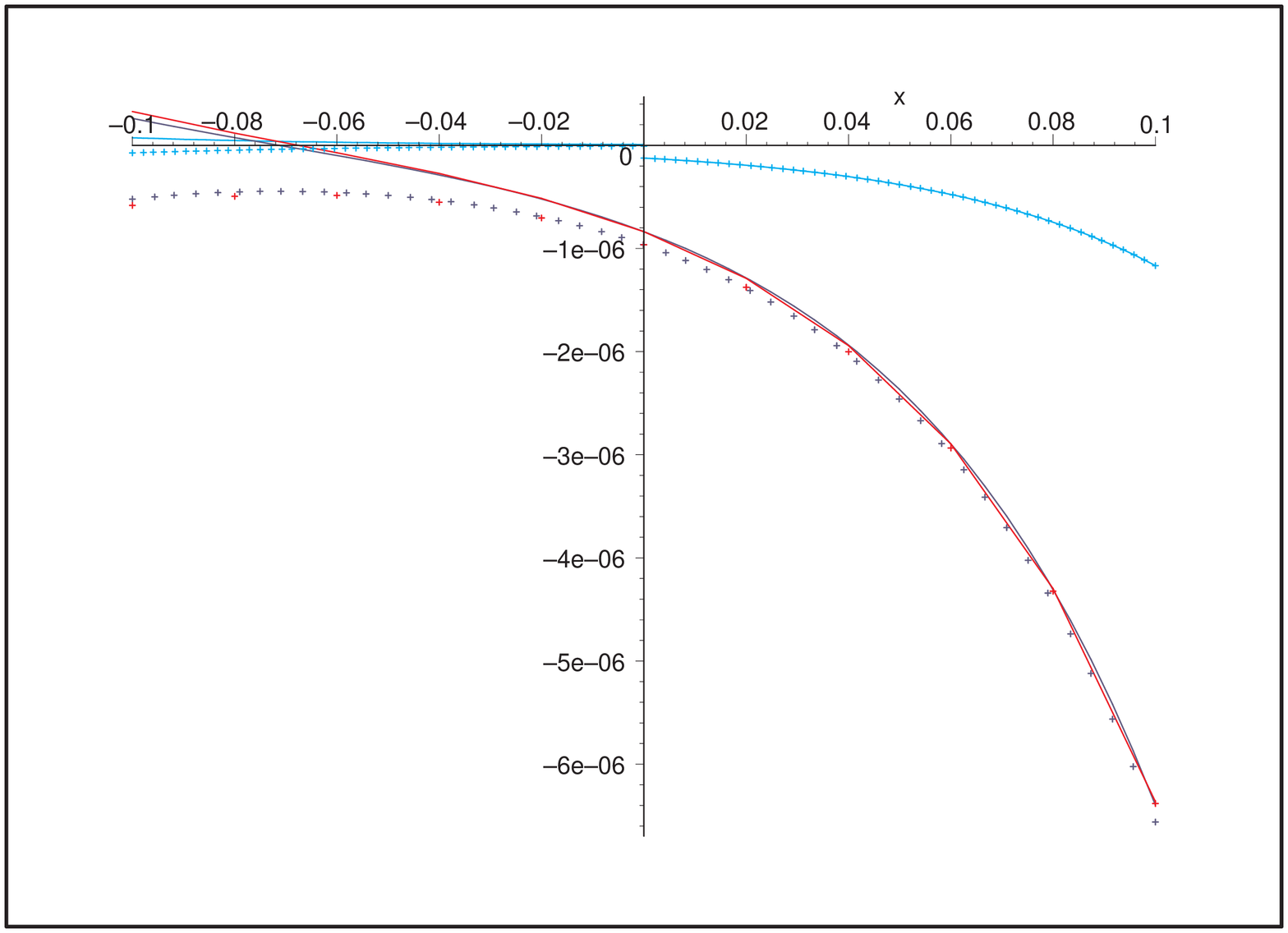}
\caption{${}_{-1}S_{l,1,25}$ for $l=3\ \&\ 4$.
The continuous lines correspond to $l=4$ and the dotted ones to $l=3$.
Correspondence between colours and solutions is the same as in Figure \ref{fig:sph_n3n4m1w5to25}.}  \label{fig:sph_n3n4m1w5to25 last}
\end{figure}

In Figures \ref{fig:sph_n3n4m1w5to25}--\ref{fig:sph_n3n4m1w5to25 last} the lines labelled as `inner' have been obtained with (\ref{eq: inner solution}),
the ones labelled `outer' with (\ref{eq: outer solution}), the ones labelled `uniform' with (\ref{eq:unif S,p,p'}) and the ones
labelled `numerics' with the programs described in Section \ref{sec:num. method; high freq. sph.}.
These figures show that the outer (normalized to agree with the numerical data at $x=0$), inner (normalized to agree with the numerical data
at $x=\pm 0.96$) and uniform (also normalized to agree with the numerical data at $x=0$) solutions
approximate the numerical data for $\omega =25$ in the boundary layers and in the neighbourhood of $x=0$. The outer solution is valid until the boundary point $x=-1$ but not
until $x=+1$ since the function has two zeros close to it and the outer solution cannot cater for them, whereas the uniform solution is a valid approximation
for all $x$. The inner solutions, on the other hand, prove to be a good approximation in the boundary layers but not close to $x=0$.

\begin{figure}[p]
\rotatebox{90}
\centering
\includegraphics*[width=90mm,angle=270]{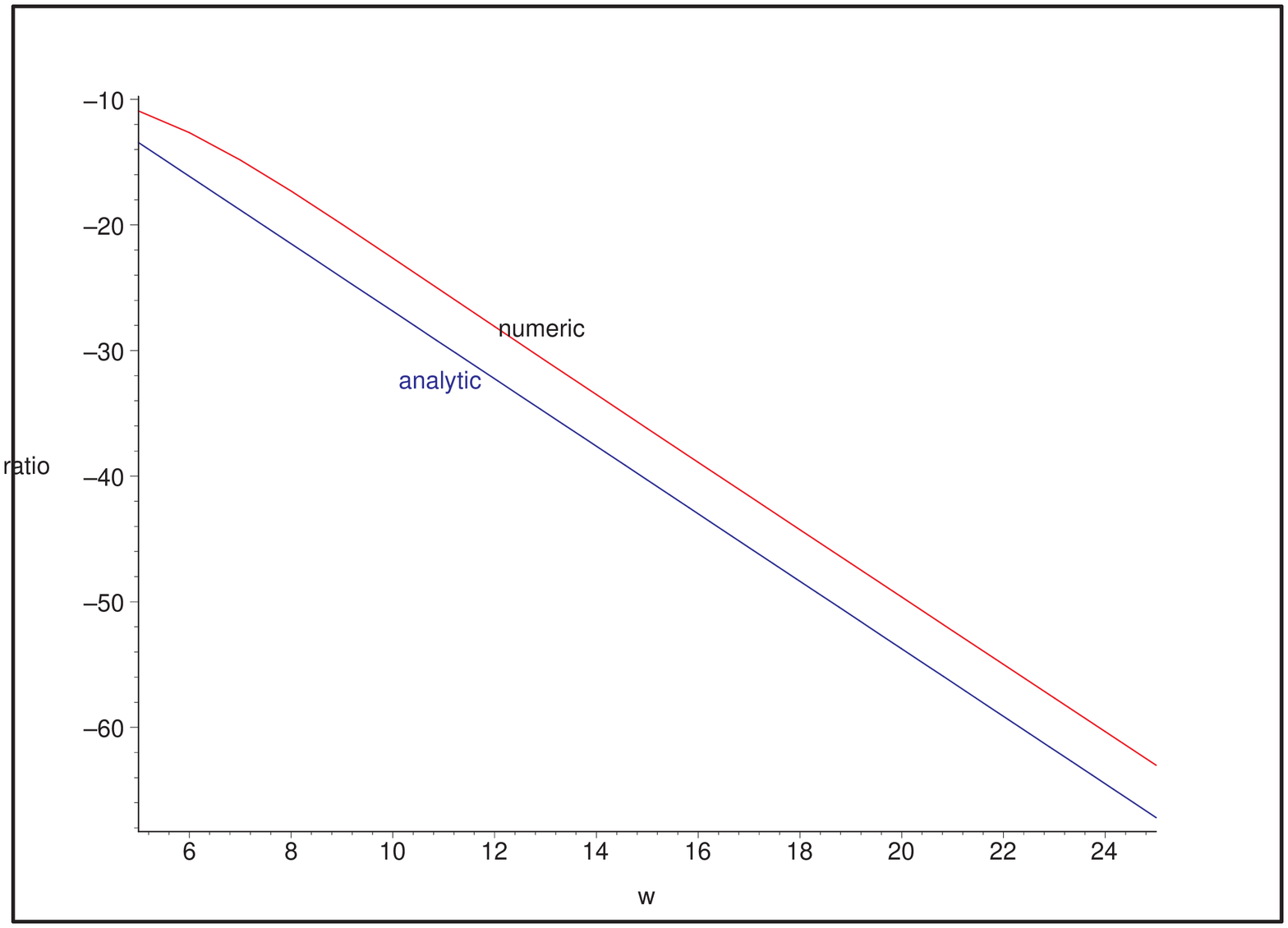}
\caption{$\frac{{}_{-1}D_{4,1,\omega}}{{}_{-1}C_{4,1,\omega}}$ for $\omega =5\rightarrow25$.
The analytic values have been obtained with (\ref{eq: ratio D/C spin1}).} \label{fig:D_1divC_1_mge1_q4n4m1w5to25}
\includegraphics*[width=90mm,angle=270]{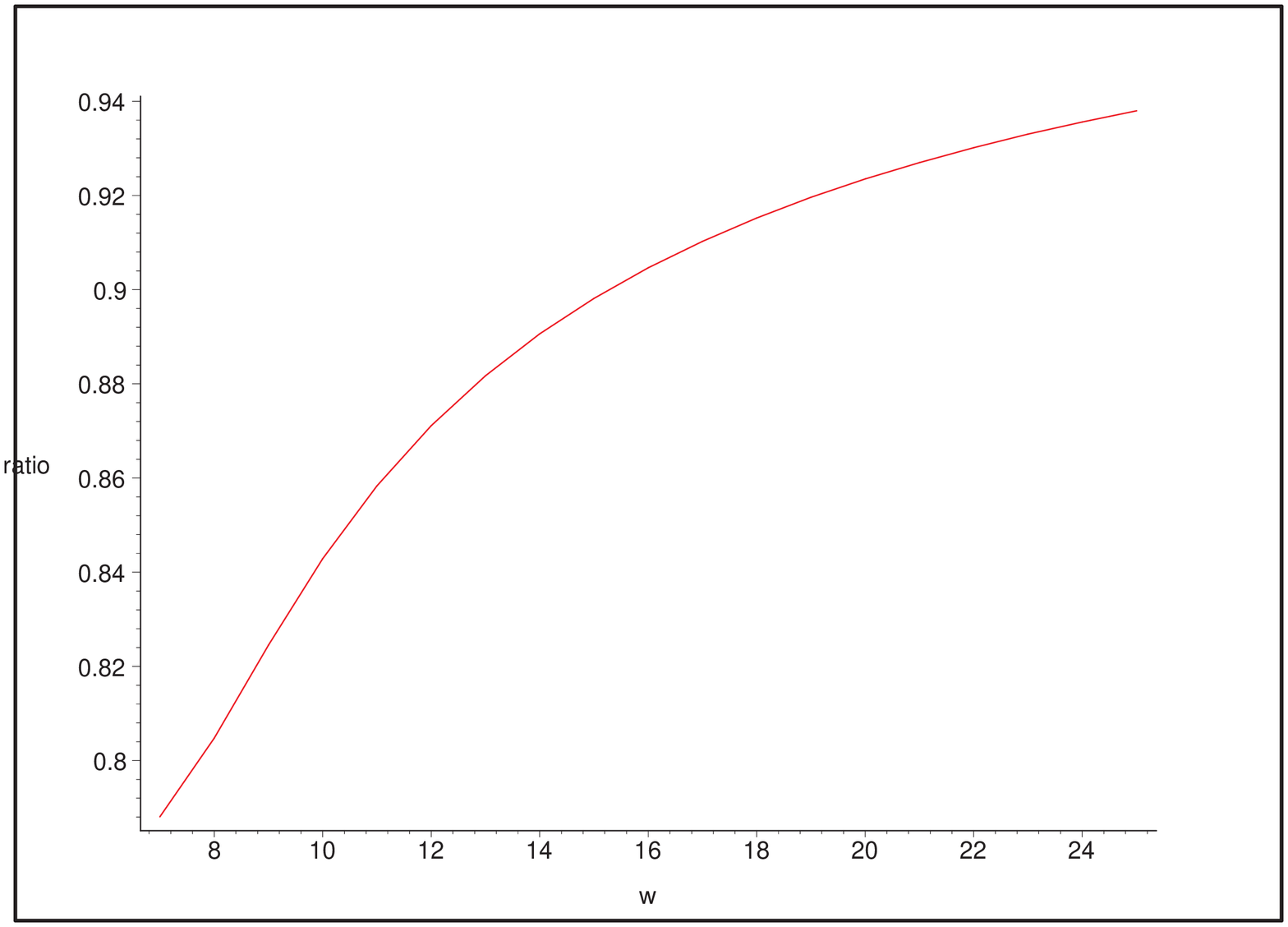}
\caption{Ratio between numeric and analytic values of $\frac{{}_{-1}D_{4,1,\omega}}{{}_{-1}C_{4,1,\omega}}$.
The analytic values have been obtained with (\ref{eq: ratio D/C spin1}).} \label{fig:ratio_D_1divC_1_mge1_q4n4m1w5to25}
\end{figure}

Figures \ref{fig:D_1divC_1_mge1_q4n4m1w5to25} and \ref{fig:ratio_D_1divC_1_mge1_q4n4m1w5to25}
prove equation (\ref{eq: ratio D/C spin1}) to be correct for the case
$m\geq 1$: for the specific values $\indhel =-1$, $l=4$, $m=1$ and $q=4$ the inner solution (\ref{eq: inner solution})
has been normalized to match the numerical data at the points $x=\pm 0.998$ for different
values of the frequency from 5 to 25, in order to be able to calculate ${}_{-1}D_{4,1,\omega}$, ${}_{-1}C_{4,1,\omega}$ and ${}_{-1}D_{4,1,\omega}/{}_{-1}C_{4,1,\omega}$.
When plotting this numerical ratio together with the analytical result (\ref{eq: ratio D/C spin1}), the two lines are parallel and therefore agree to
highest order, and the ratio between the numerical and the analytical data tends to $1$.

\begin{figure}[p]
\rotatebox{90}
\centering
\includegraphics*[width=90mm,angle=270]{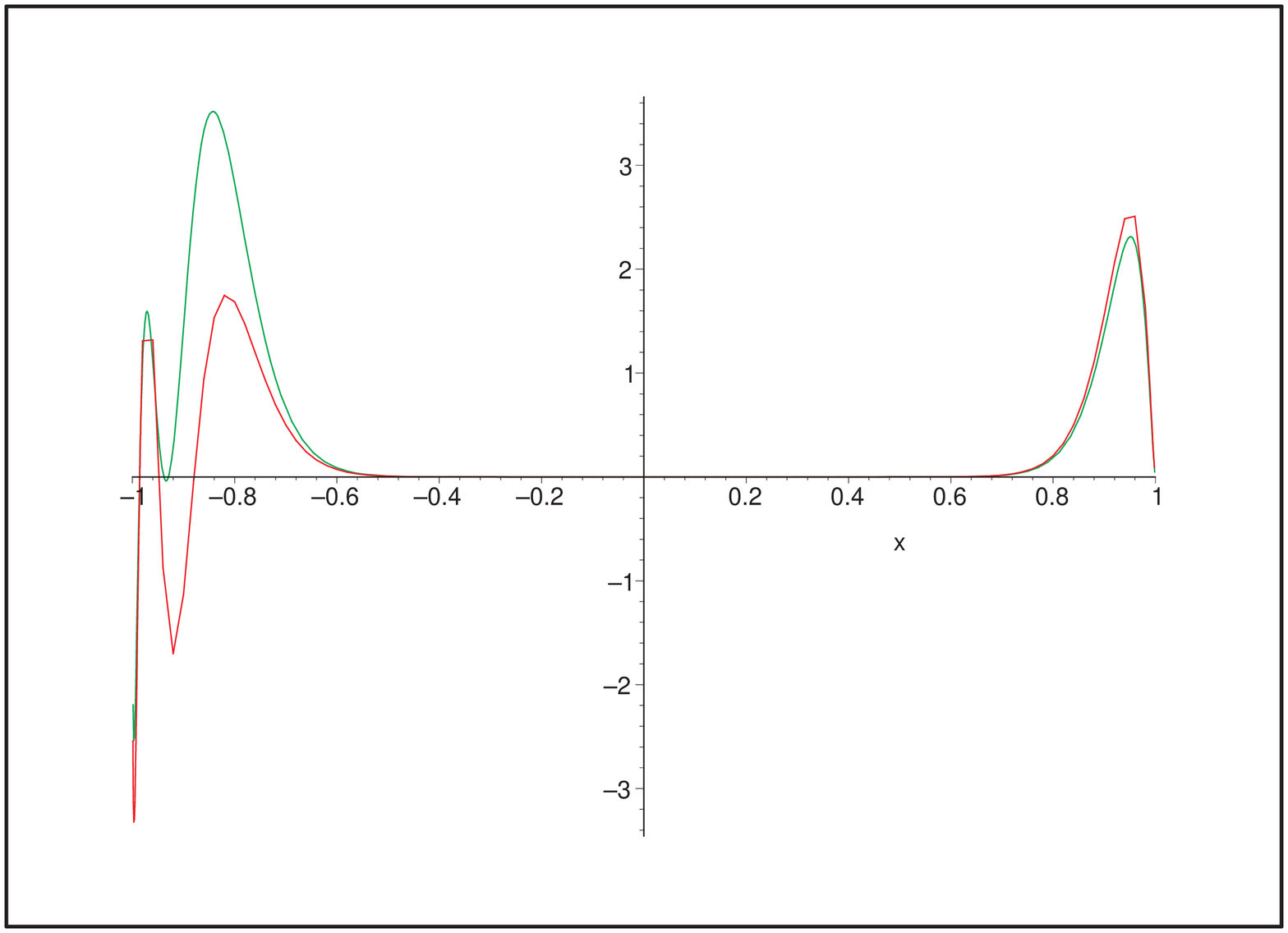}
\caption{${}_{+2}S_{5,1,35}$.
Green line corresponds to uniform solution (\ref{eq:unif S,p,p'}) and red line to numerics.} \label{fig:sph_s2n5m1w1to37_w35_x_1to1}
\includegraphics*[width=90mm,angle=270]{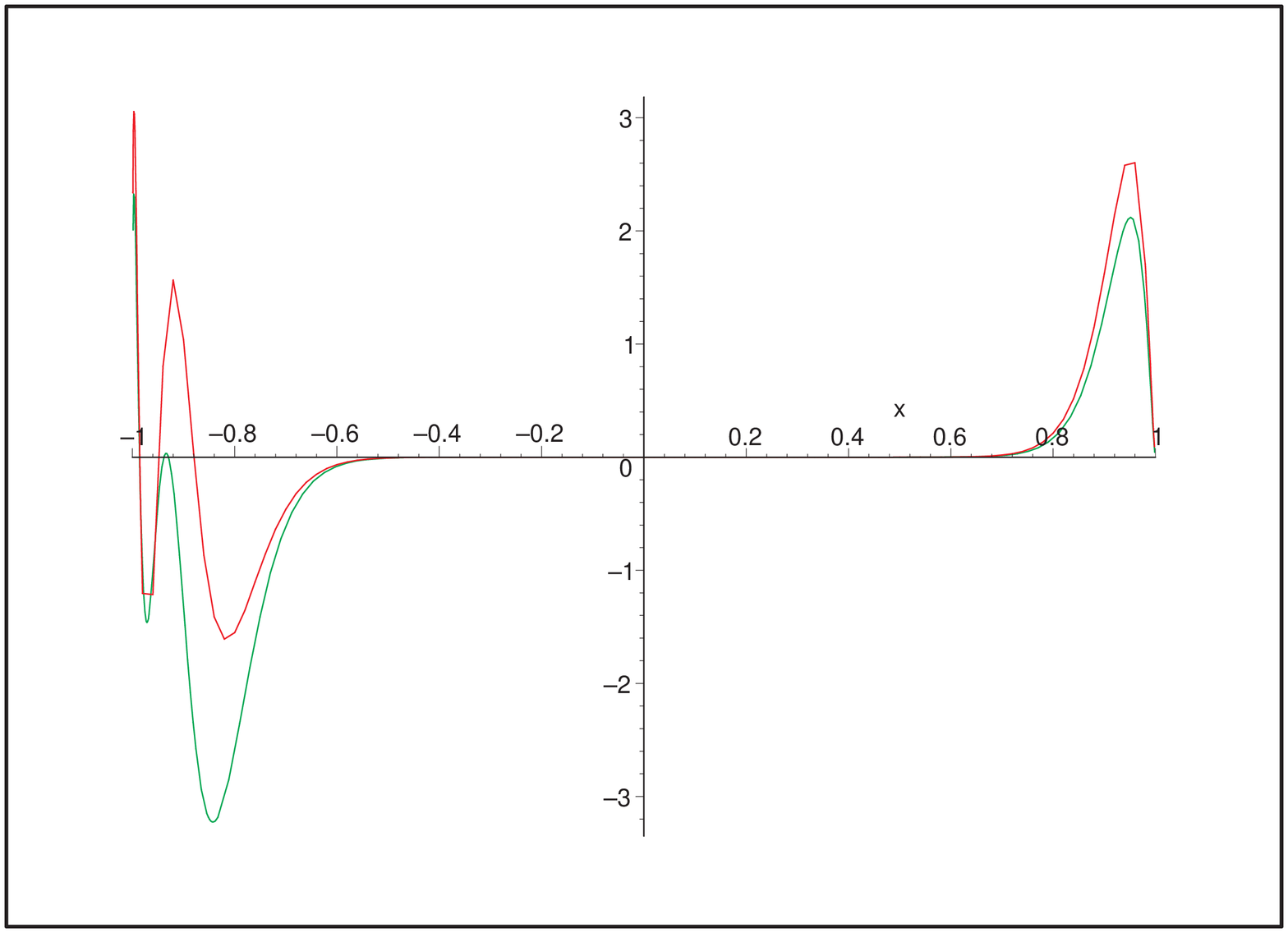}
\caption{${}_{+2}S_{6,1,35}$.
Green line corresponds to uniform solution (\ref{eq:unif S,p,p'}) and red line to numerics.}
\label{fig:sph_s2n5n6m1w1to35}
\end{figure}

\begin{figure}[!p]
\rotatebox{90}
\centering
\includegraphics*[width=100mm,angle=270]{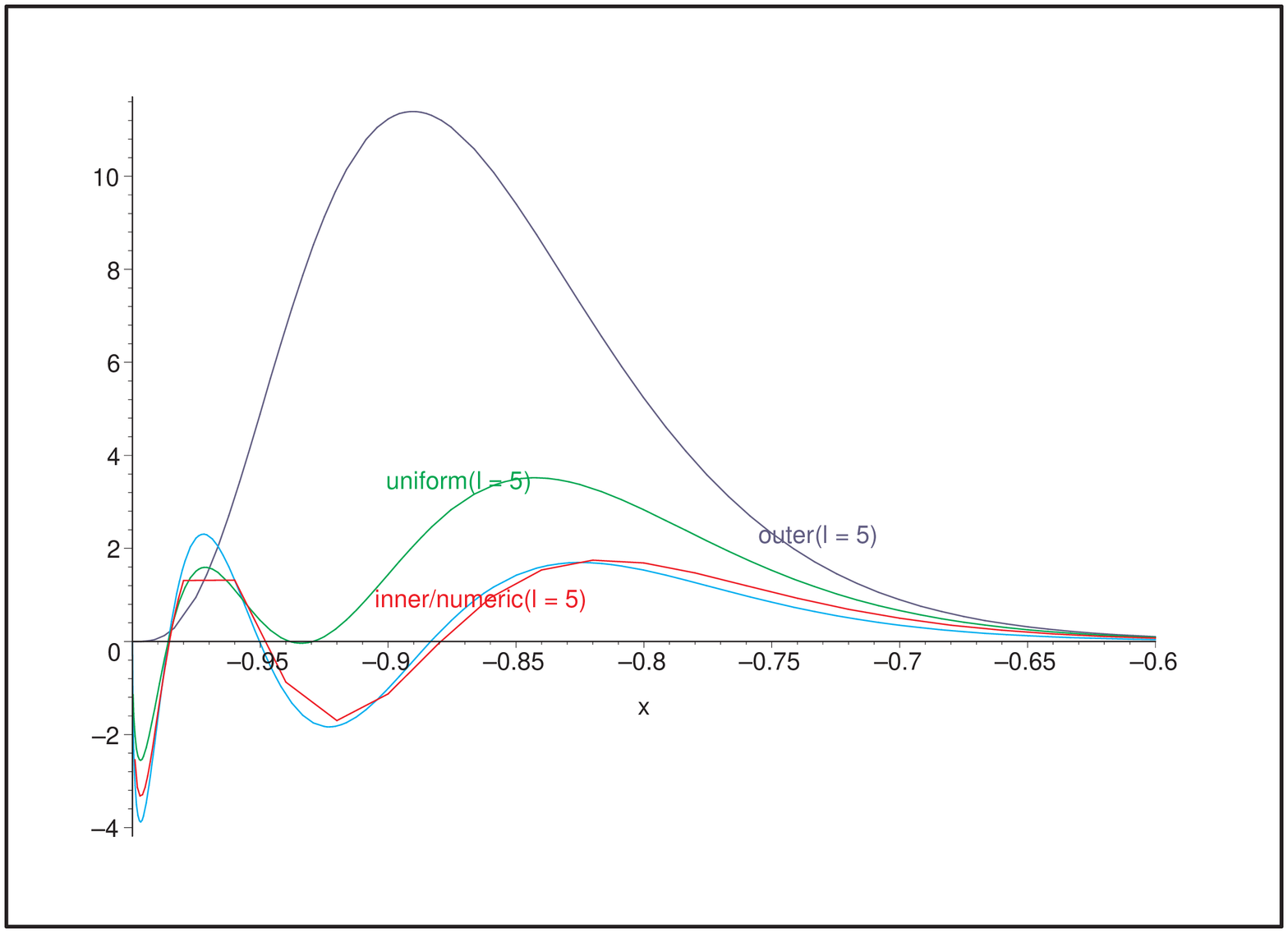}
\caption{${}_{+2}S_{5,1,35}$. Correspondence between colours and solutions is the same as in Figure \ref{fig:sph_n3n4m1w5to25}.}
\label{fig:sph_s2n5m1w1to37_w35_x_0p6to_1}
\includegraphics*[width=100mm,angle=270]{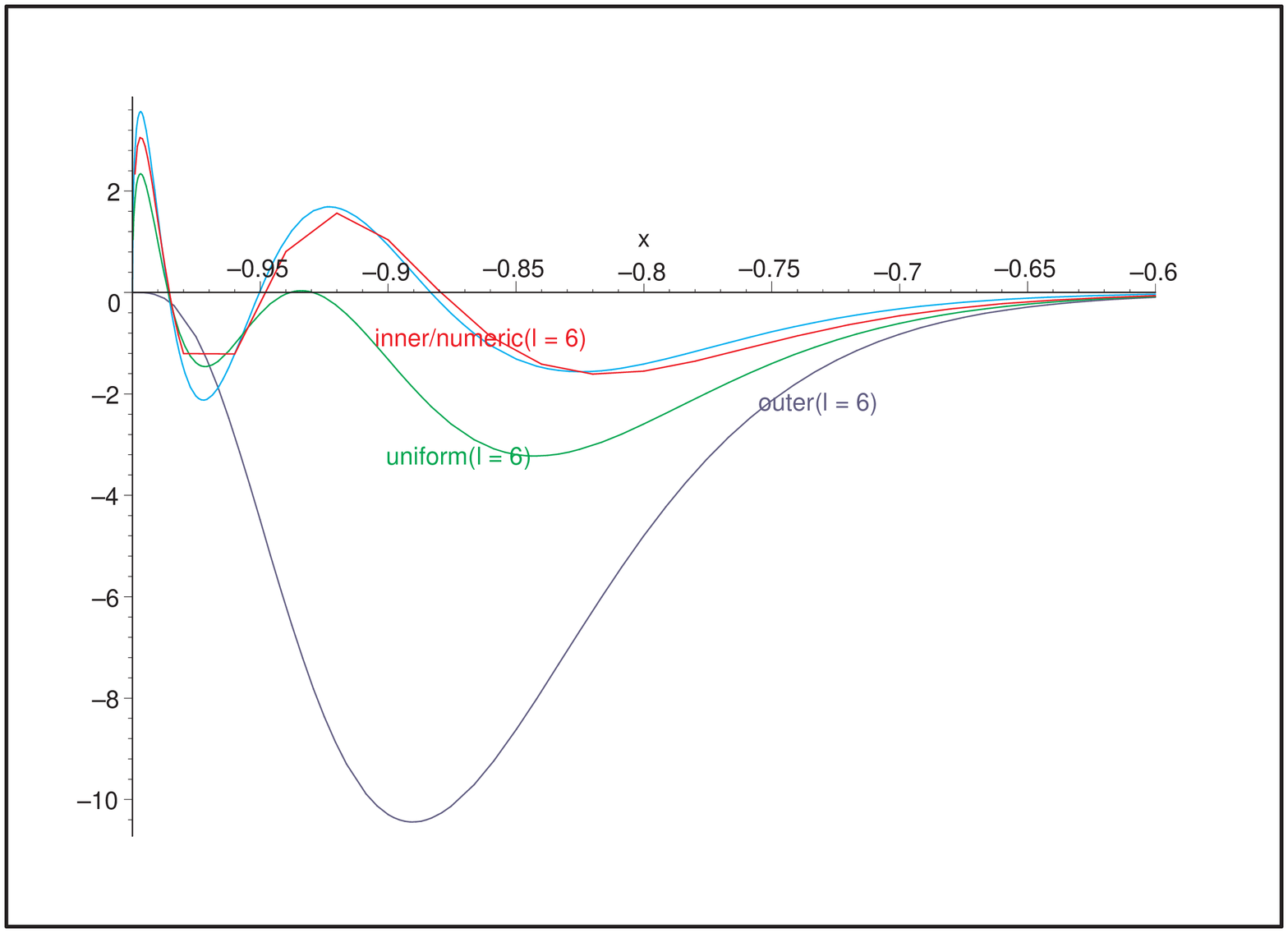}
\caption{${}_{+2}S_{6,1,35}$. Correspondence between colours and solutions is the same as in Figure \ref{fig:sph_n3n4m1w5to25}.}
\label{fig:sph_s2n6m1w1to37_w35_x_0p6to_1}
\end{figure}

Figures \ref{fig:sph_s2n5m1w1to37_w35_x_1to1}--\ref{fig:sph_s2n6m1w1to37_w35_x_0p6to_1}
correspond to modes with $h=+2$, $m=1$, $\omega=35$ and $l=5$ or $l=6$. The modes for both values of $l$ yield $q=6$.
However, the mode with $l=5$ does not possess a zero at $x_0$ whereas the mode with $l=6$ does.
The behaviour for positive $x$ is very similar for both values of $l$ but for negative $x$ the behaviours for the two modes differ by a sign.

\begin{figure}[p]
\rotatebox{90}
\centering
\includegraphics*[width=90mm,angle=270]{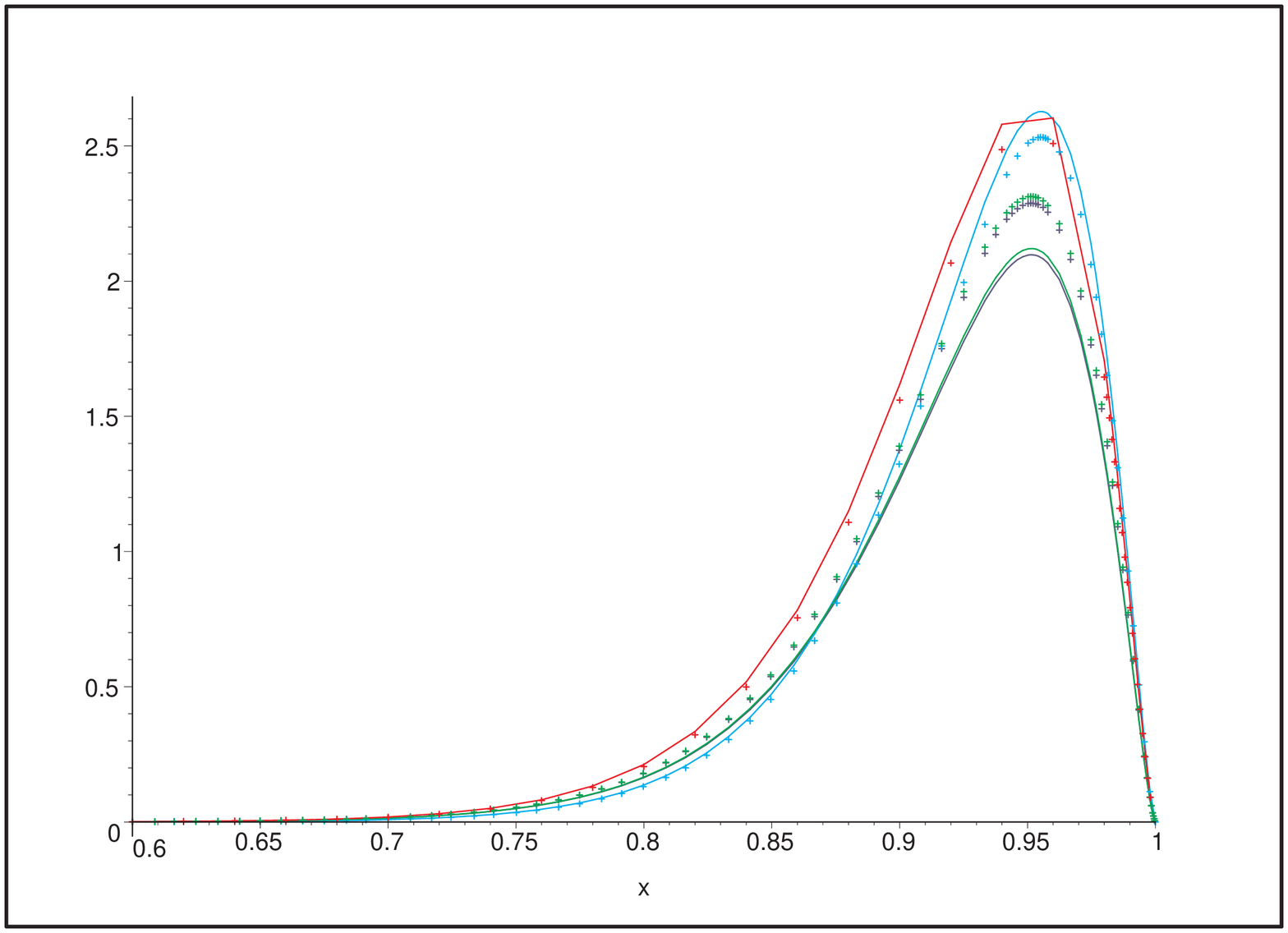}
\caption{${}_{+2}S_{l,1,35}$ for $l=5\ \&\ 6$.
The continuous lines correspond to $l=6$ and the dotted ones to $l=5$.
Correspondence between colours and solutions is the same as in Figure \ref{fig:sph_n3n4m1w5to25}.}
\includegraphics*[width=90mm,angle=270]{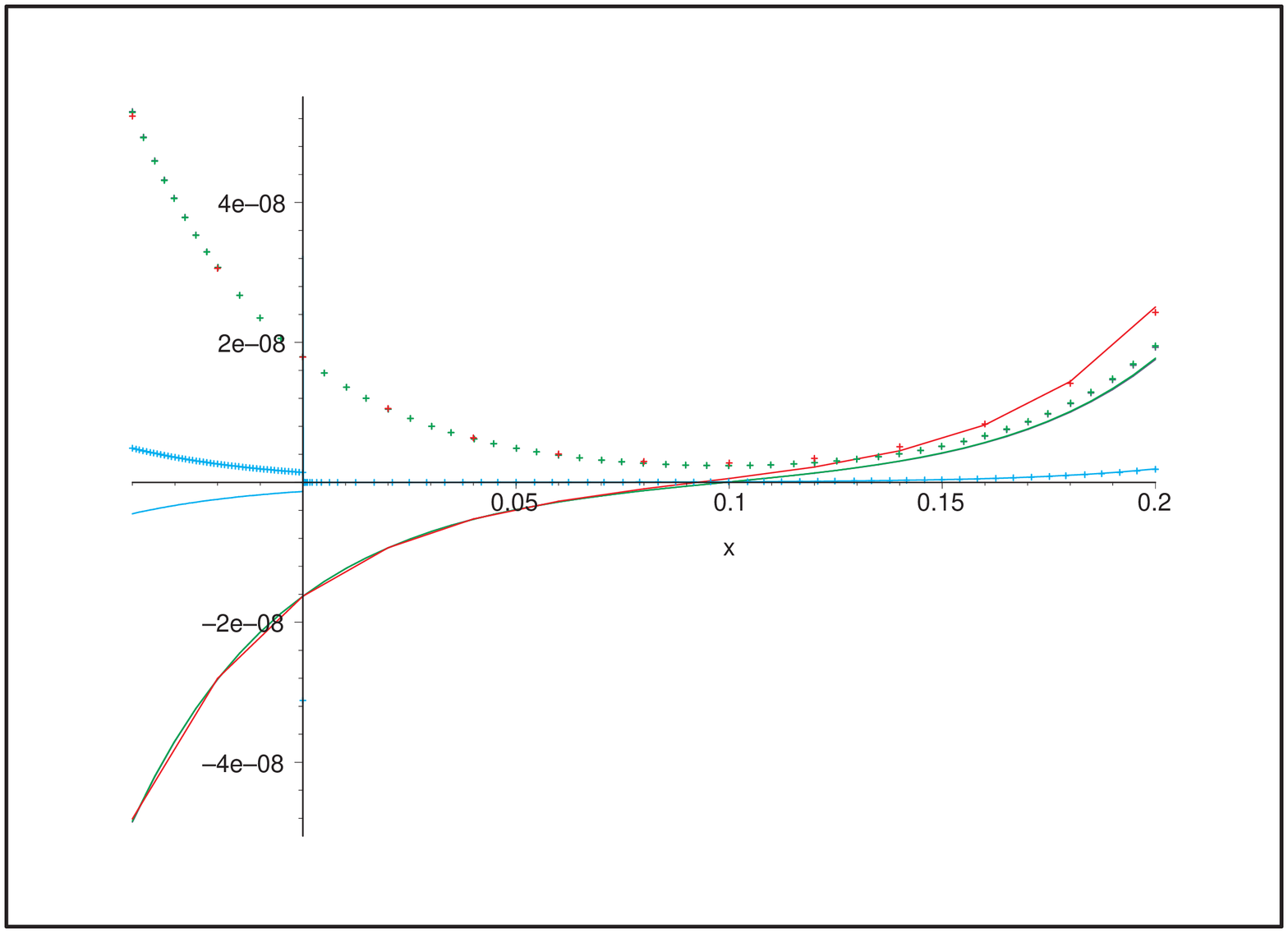}
\caption{${}_{+2}S_{l,1,35}$ for $l=5\ \&\ 6$.
The continuous lines correspond to $l=6$ and the dotted ones to $l=5$.
Correspondence between colours and solutions is the same as in Figure \ref{fig:sph_n3n4m1w5to25}.}
\end{figure}


\begin{figure}[!ht]
\rotatebox{90}
\centering
\includegraphics*[width=90mm,angle=270]{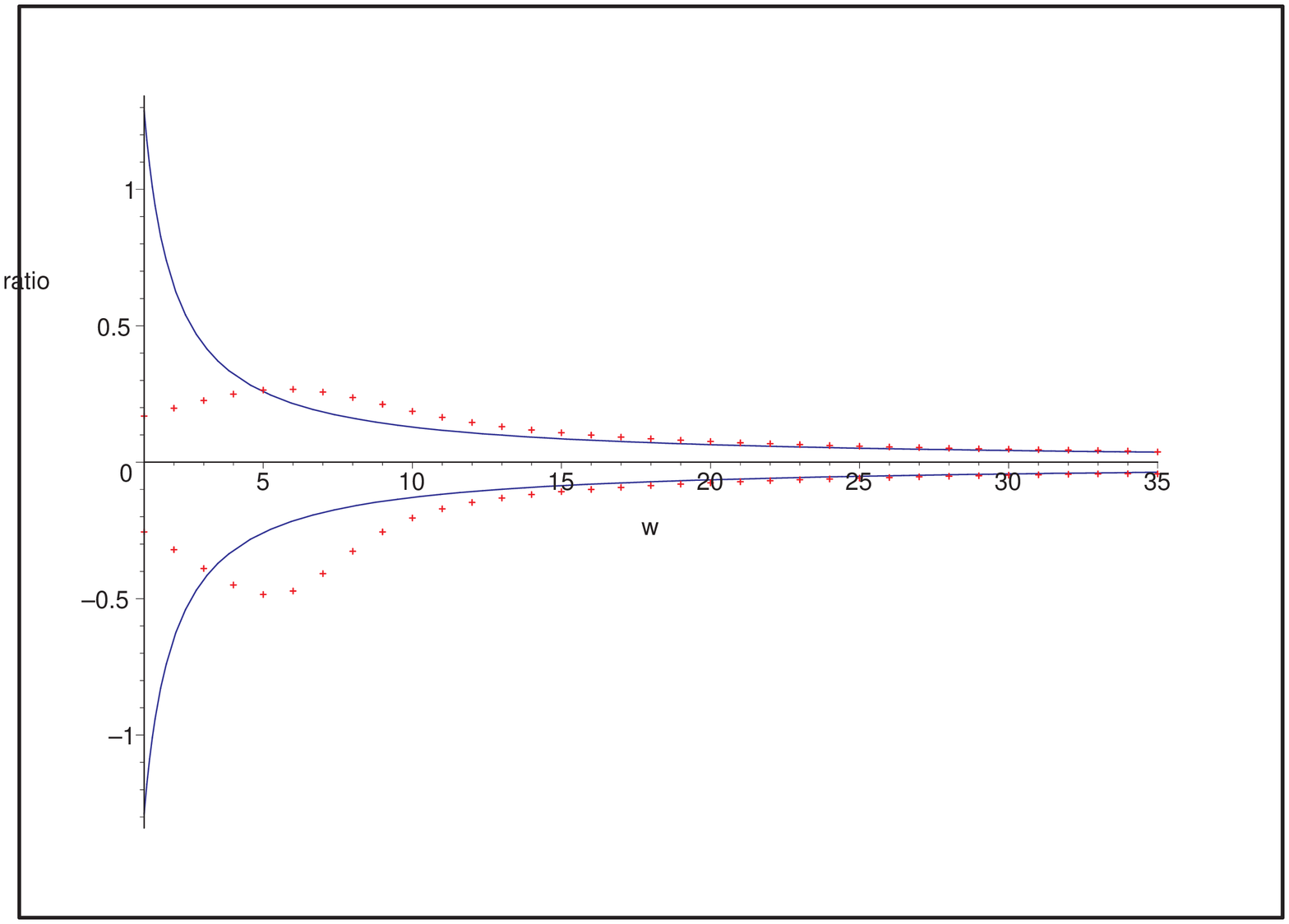}
\caption{$\frac{{}_{-1}D_{4,1,\omega}}{{}_{-1}C_{4,1,\omega}}$ for $\omega =1\rightarrow 35$.
The curves above the x-axis correspond to $l=6$ and below to $l=5$.
The continuous lines correspond to the analytic expression (\ref{eq: ratio D/C spin2}) and the dotted ones to the numerical data.}
\includegraphics*[width=90mm,angle=270]{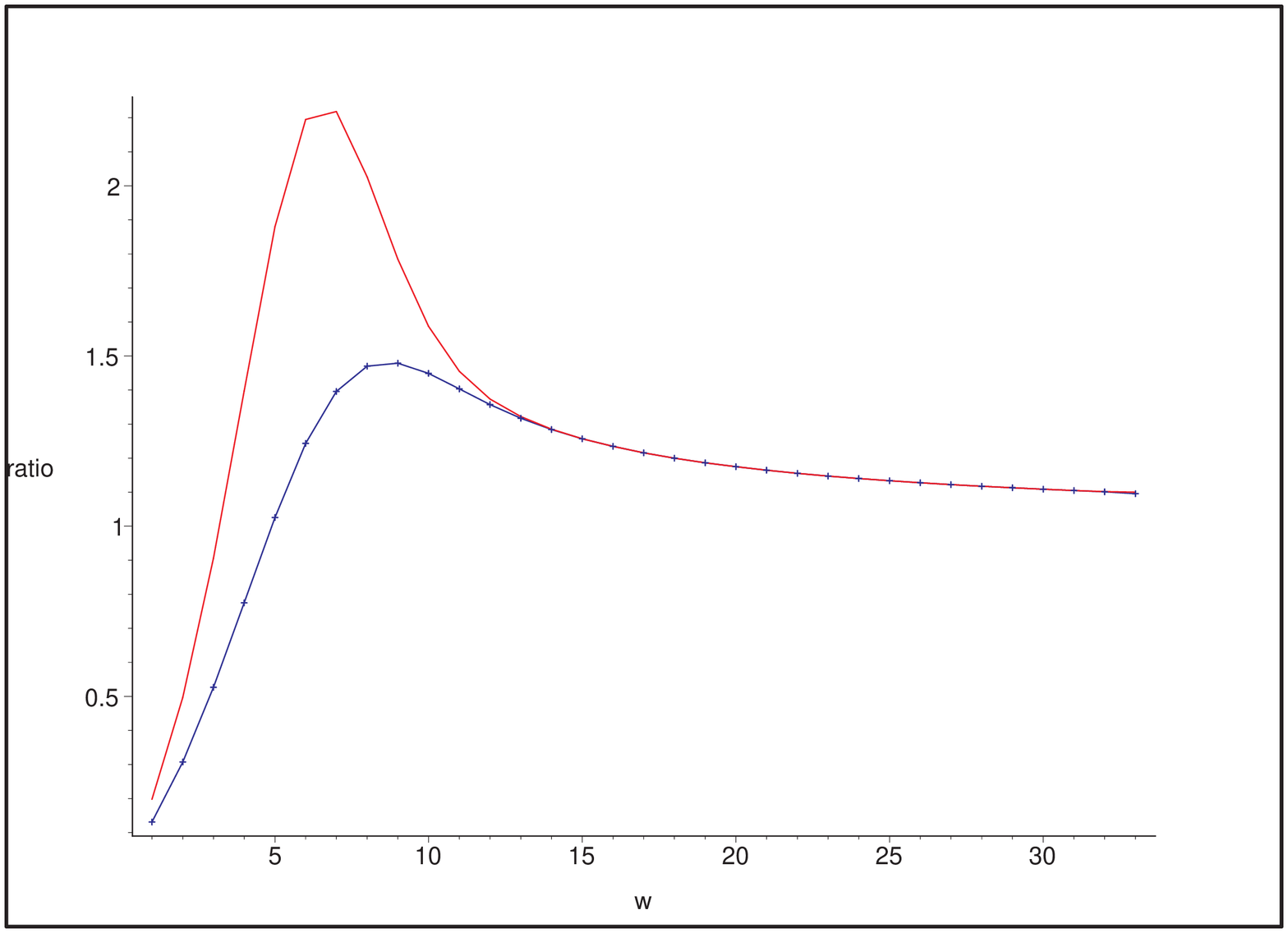}
\caption{Ratio between numeric and analytic values of $\frac{{}_{+2}D_{l,1,\omega}}{{}_{+2}C_{l,1,\omega}}$ for $\omega =1\rightarrow 34$.
Blue lines (plotted both continuous and dotted to show agreement with red line for large $\omega$) correspond to $l=6$ and
red line to $l=5$.
} \label{fig:ratio_D2divC2_m1_s2q6n5n6m1w1to34_x0p998}
\end{figure}

\begin{figure}[p]
\rotatebox{90}
\centering
\includegraphics*[width=90mm,angle=270]{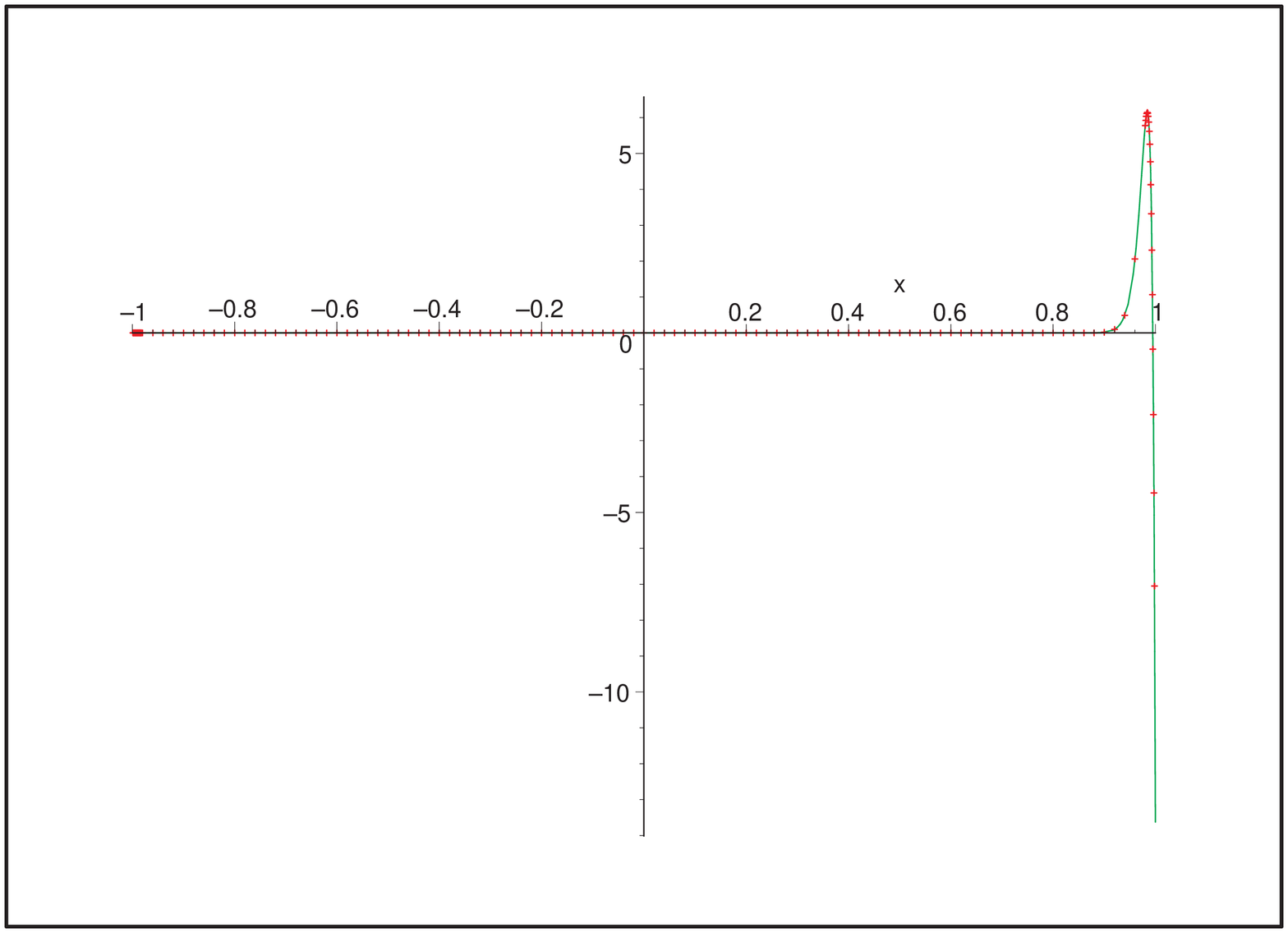}
\caption{${}_{-1}S_{2,1,100}$.
The continuous, green line corresponds to the uniform solution (\ref{eq:unif S,p,p' not}) and the dotted, red one to the numerical data}
\includegraphics*[width=90mm,angle=270]{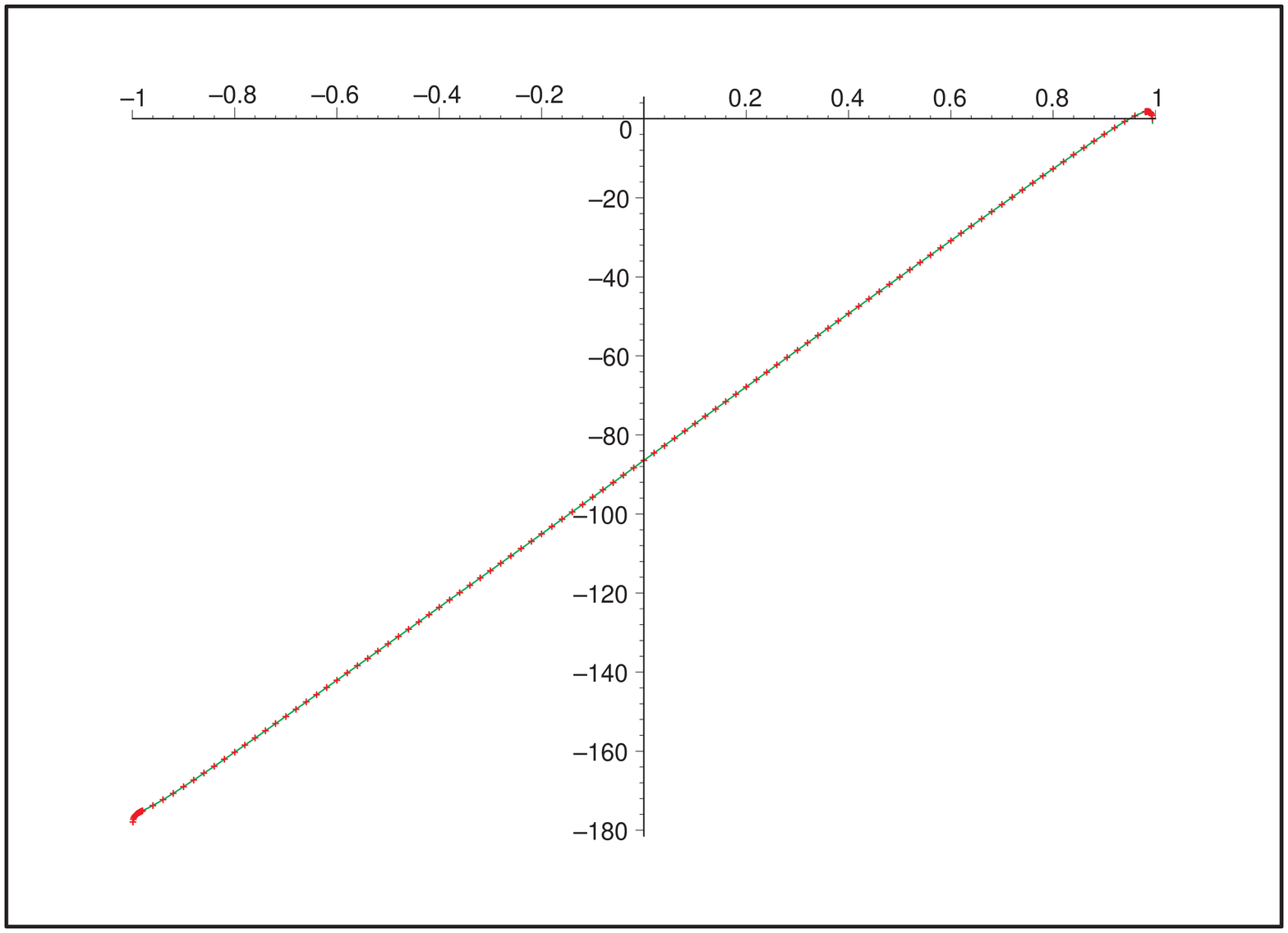} \label{fig:sph_n2m1w100_x_1to1}
\caption{$\log({}_{-1}S_{2,1,100})$.
The continuous, green line corresponds to the uniform solution (\ref{eq:unif S,p,p' not}) and the dotted, red one to the numerical data}
\label{fig:sph_n2m1w100}
\end{figure}

\begin{figure}[p]
\rotatebox{90}
\centering
\includegraphics*[width=90mm,angle=270]{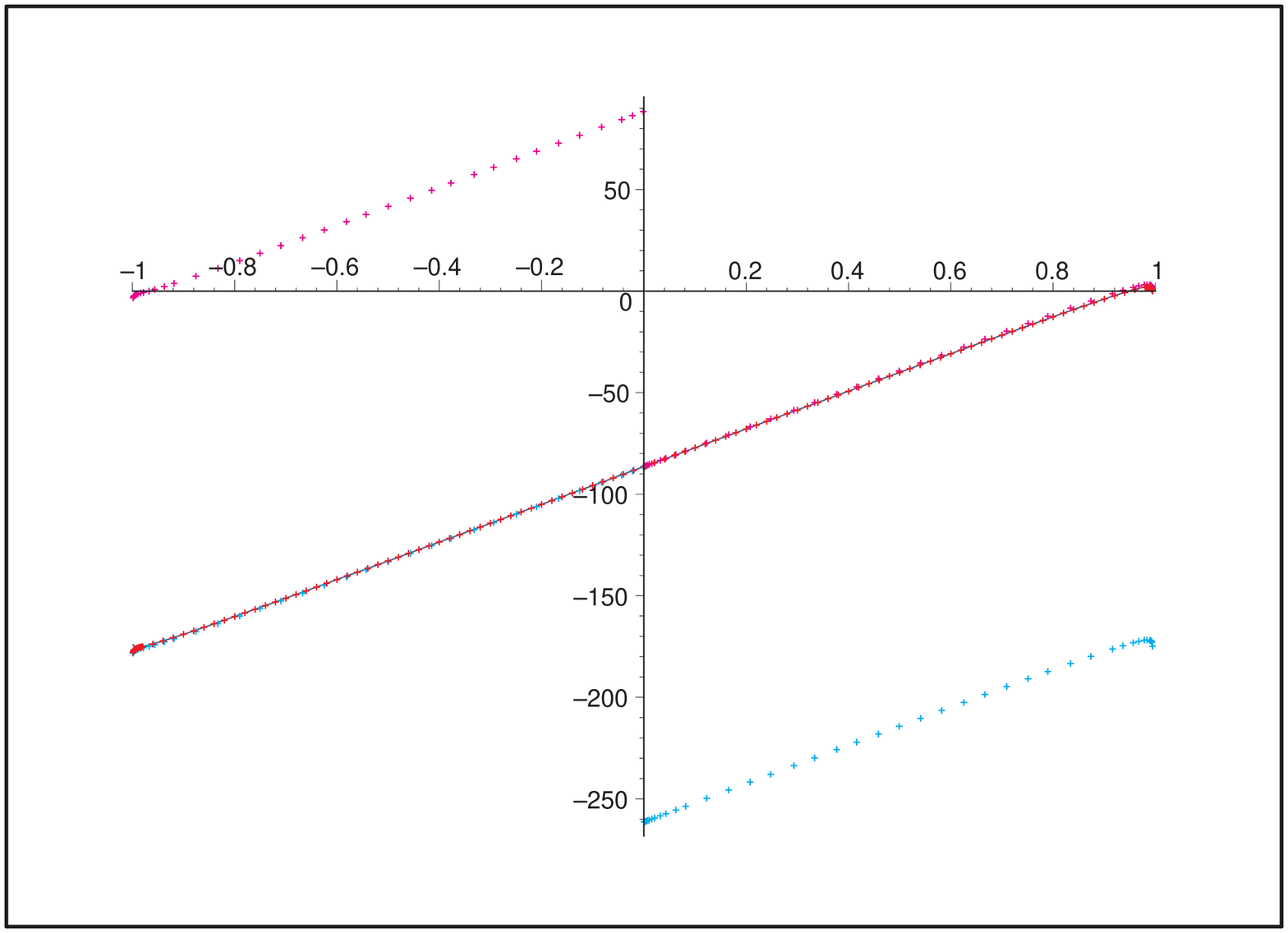}
\caption{$\log({}_{-1}S_{2,1,100})$.
The red line (numerical data) overlaps with the navy line (outer solution).
The light blue line (inner solution valid at  $x\sim -1$)
and the magenta line (inner solution valid at  $x\sim +1$) overlap with the red/navy lines for negative and positive $x$ respectively.}
\label{fig:log_sph_n2m1w100_x_1to1}
\includegraphics*[width=90mm,angle=270]{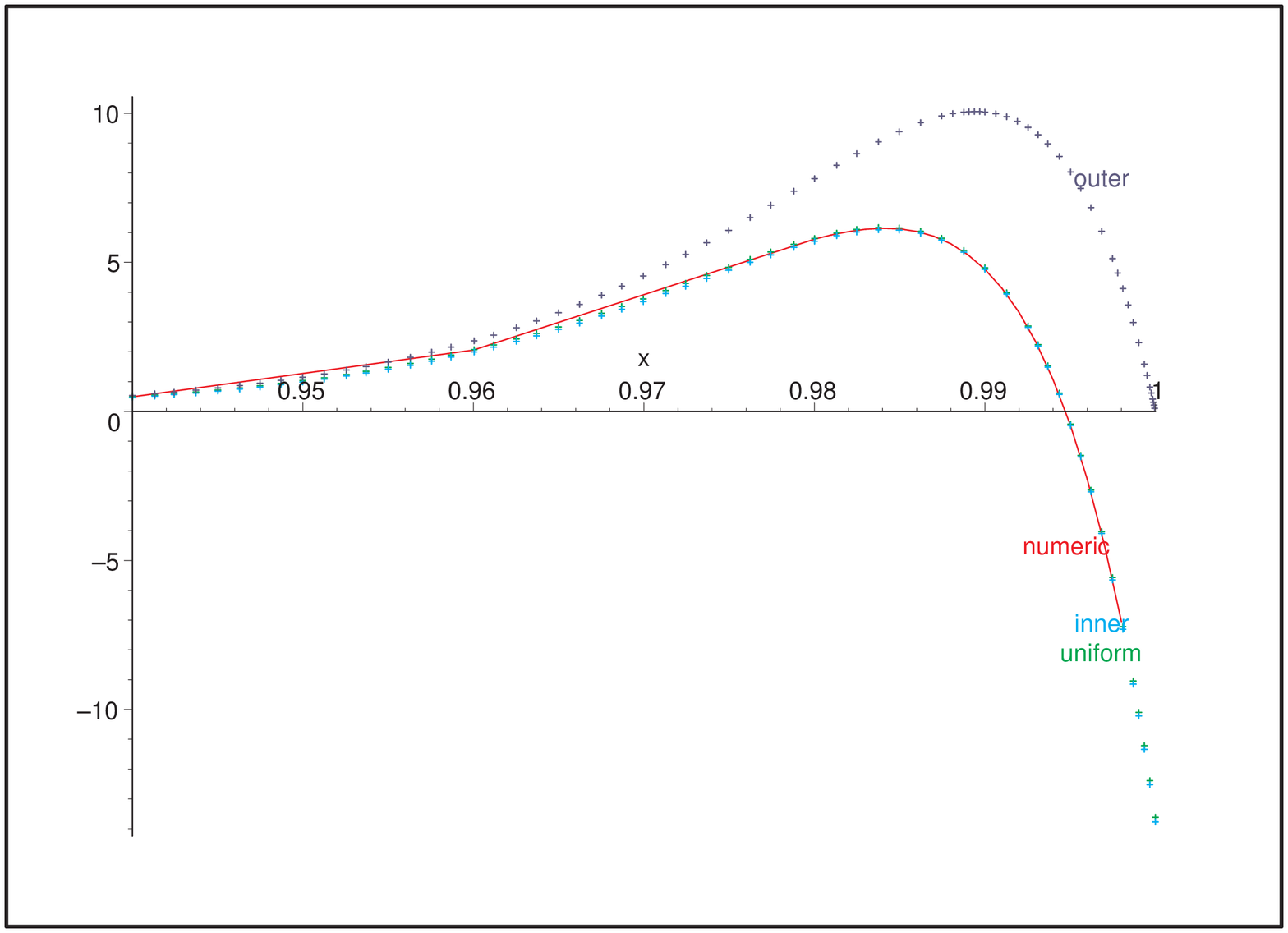}
\caption{${}_{-1}S_{2,1,100}$.} \label{fig:sph_unif_n2m1w100_x0p94to1}
\end{figure}

For $\indhel =-1$, $l=2$, $m=1$ and $\omega =100$, the corresponding value of $q$ is $2$.
This is a case where $p\in \mathbb{Z}^{+}\cup\{0\}$ and
$p'\notin \mathbb{Z}^{+}\cup\{0\}$. The numerical solution together with the uniform expansion (\ref{eq:unif S,p,p' not}) is plotted over the whole range
$x\in[-1,1]$ in Figures \ref{fig:sph_n2m1w100_x_1to1}--\ref{fig:log_sph_n2m1w100_x_1to1}.

As we have seen, in this case the function has an exponential
behaviour far from the boundary layers, so that a plot of the
$\log$ of the function allows us to see the behaviour over the
whole range of $x$. Both the uniform expansion and the outer
solution have been normalized so that they coincide with the
numerical value at $x=0$, and the inner solution has been
normalized once at $x=10^{-8}$ and once at $x=-10^{-8}$. The
uniform expansion agrees with the numerical solution for all
values of $x$. The outer solution agrees with the numerics
everywhere except very close to $x=\pm 1$, where it veers off.
The inner solutions are valid all the way from their respective
boundary layers until, and past, $x=0$, which is due to the
exponential nature of the function in the region between the
boundary layers. The inner solutions show a jump at $x=0$ due to
the different orders in $c$ of ${}_{\indhel}C_{lm\omega}$ and ${}_{\indhel}D_{lm\omega}$.

\begin{figure}[p]
\rotatebox{90}
\centering
\includegraphics*[width=90mm,angle=270]{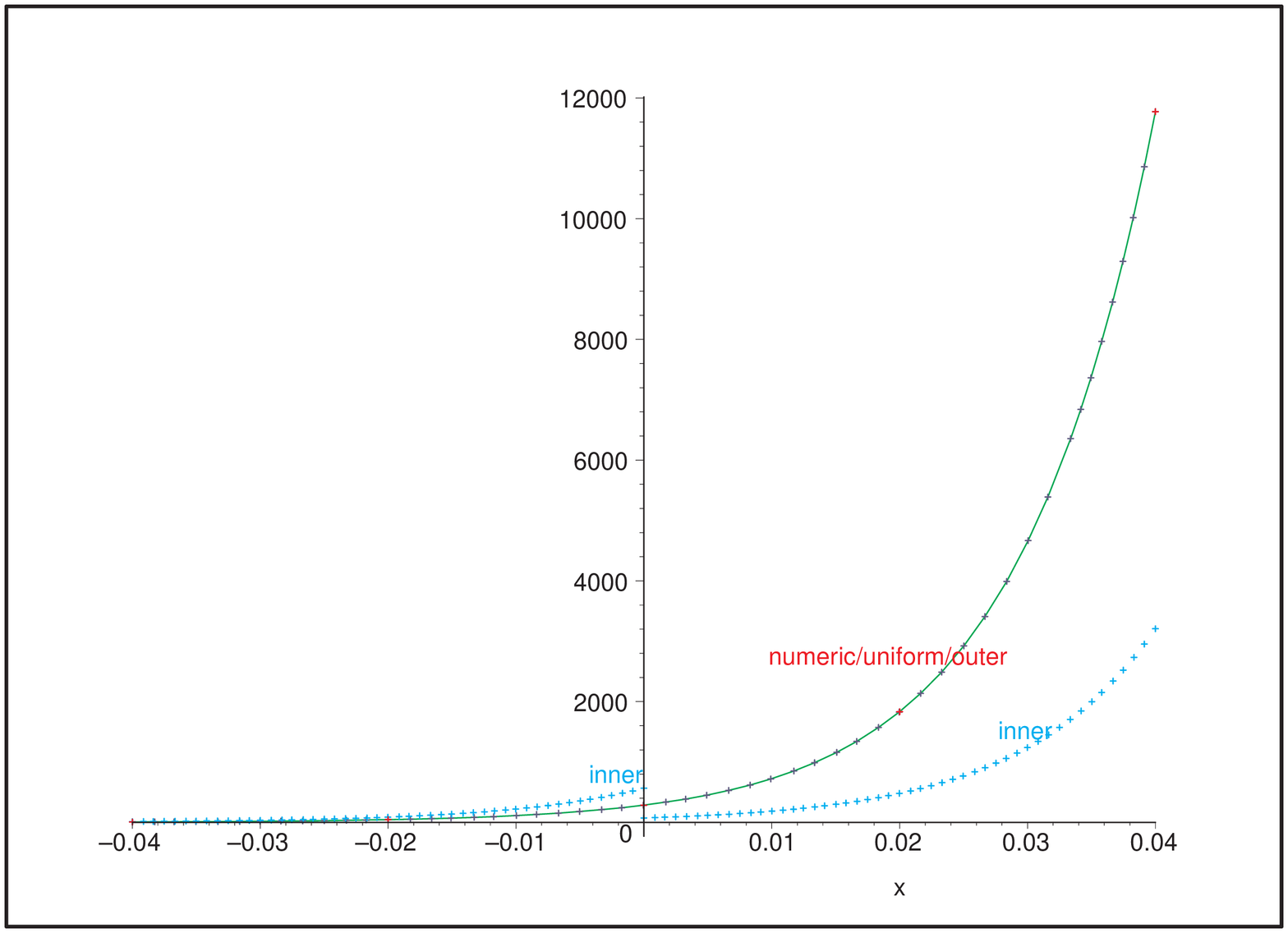}
\caption{$10^{40}{}_{-1}S_{2,1,100}$}\label{fig:sph_unif_n2m1w100_x_0p04to0p04}
\includegraphics*[width=90mm,angle=270]{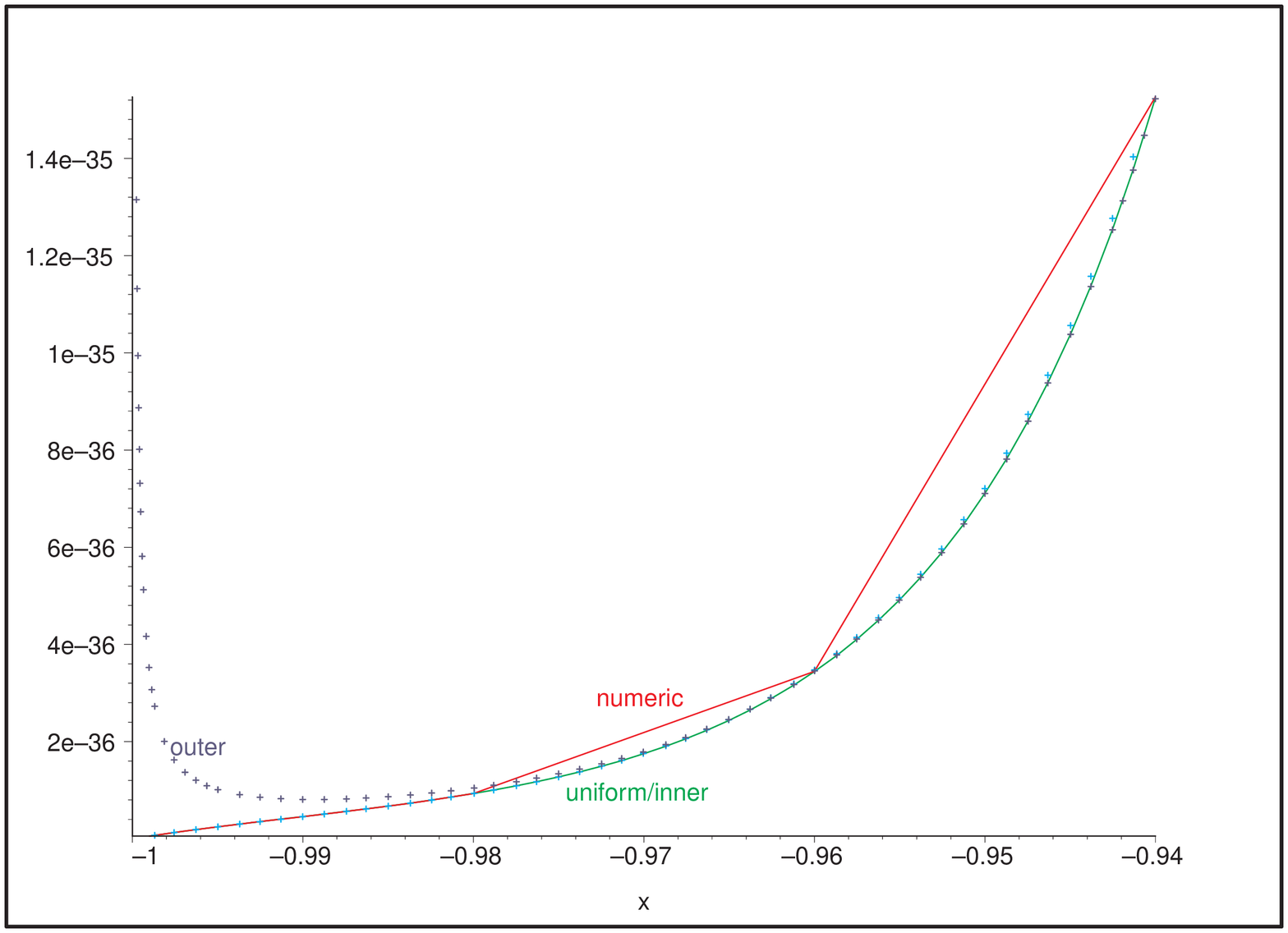}
\caption{$10^{40}{}_{-1}S_{2,1,100}$} \label{fig:sph_unif_n2m1w100_x_0p04to_1}
\end{figure}

The above features can be seen in detail for $x$ close to $0$ and $\pm 1$ in
Figures \ref{fig:sph_unif_n2m1w100_x0p94to1}--\ref{fig:sph_unif_n2m1w100_x_0p04to_1}
where they have been rescaled by $10^{40}$ for $x$ close to $0$ and $-1$.

\newpage


\section*{Acknowledgements}

We wish to thank Ted Cox for his helpful contribution. 
We also wish to thank Enterprise Ireland for financial support.


\bibliography{./papers,./books,./refs}

\end{document}